\documentclass[aps,prl,twocolumn,superscriptaddress,nobibnotes,amsmath,amssymb]{revtex4-2}
\usepackage{graphicx}
\usepackage{dcolumn}
\usepackage{bm}
\usepackage{natbib}
\usepackage{color}
\usepackage{amsmath}
\usepackage[]{hyperref}
\hypersetup{
    colorlinks=true,
    allcolors = {blue},
    }
\usepackage[normalem]{ulem}
\newcommand{\remove}[1]{}

\newcommand{\add}[1]{\textcolor{black}{#1}}
\begin{document}

\title{Thermal bottleneck in a freely suspended superconducting island on InAs nanowire}
\author{E.V.~Shpagina}\affiliation{Osipyan Institute of Solid State Physics, Russian Academy of
Sciences, 142432 Chernogolovka, Russian Federation}
\affiliation{National Research University Higher School of Economics, 20 Myasnitskaya Street, 101000 Moscow, Russian Federation}
\author{E.S.~Tikhonov}\affiliation{Osipyan Institute of Solid State Physics, Russian Academy of
Sciences, 142432 Chernogolovka, Russian Federation}
\affiliation{National Research University Higher School of Economics, 20 Myasnitskaya Street, 101000 Moscow, Russian Federation}
\author{D.~Ruhstorfer}
\affiliation{Walter Schottky Institut, Physik Department, and Center for Nanotechnology and Nanomaterials, Technische Universit\"{a}t M\"{u}nchen, Am Coulombwall 4, Garching 85748, Germany}
\author{G.~Koblm\"{u}ller}
\affiliation{Walter Schottky Institut, Physik Department, and Center for Nanotechnology and Nanomaterials, Technische Universit\"{a}t M\"{u}nchen, Am Coulombwall 4, Garching 85748, Germany}
\author{V.S.~Khrapai}
\affiliation{Osipyan Institute of Solid State Physics, Russian Academy of
Sciences, 142432 Chernogolovka, Russian Federation}
\affiliation{National Research University Higher School of Economics, 20 Myasnitskaya Street, 101000 Moscow, Russian Federation}
\email{dick@issp.ac.ru}

\begin{abstract} 
We investigate the heat balance in superconducting islands (S-islands) formed in epitaxial Al/InAs nanowires (NWs) freely suspended above the substrate. We employ a Joule spectroscopy approach, which traces the superconductor-normal transition in the S-island mediated by heating of the neighboring InAs NW segments via transport current. The temperature of the surrounding $^3\mathrm{He}$ bath is varied with nearby mesoscopic heaters and controlled with the NW Johnson noise thermometry. The experiment reveals a substantial thermal relaxation bottleneck associated with the cooling via surrounding $^3\mathrm{He}$, which gives rise to phonon heating in the S-island. Our results uncover the role of environmental cooling in non-equilibrium experiments in S-islands in NW devices.
\end{abstract}

\maketitle

Recently, hybrid semiconductor nanowire (NW) -- superconductor (S) devices attracted interest for non-equilibrium thermal transport. In the context of Majorana research, a low energy electronic thermal conductance can serve as a direct marker of the topological phase transition~\cite{akhmerov2011,pan2021,bubis2021}, however, experiments of this kind are difficult and therefore scarce~\cite{denisov2021, denisov2022, majidi2022, haldar2026}. The main focus  of the non-equilibrium experiments in hybrid NWs~\cite{ibabe2023, liu2023, shpagina2024, ibabe2024,shpagina2025,oliveira2025} is the behavior of superconducting order parameter in the superconducting shell upon the injection of hot quasiparticles from the semiconducting core.  Similar experiments were recently perfromed in bilayer graphene-superconductor hybrids~\cite{galante-agero2026}. In a superconducting island (S-island) on the NW the  quasiparticles can be trapped for timescales much longer than the inelastic relaxation time, such that the local equilibrium is almost reached and the superconducting order parameter follows the local electron temperature ($T_\mathrm{e}$).

The simplest two-temperature model (2TM)~\citep{courtois2008,maisi2013} relates the $T_\mathrm{e}$, the total Joule heat $P_\mathrm{J}$ released in the semiconducting core of the NW and the bath temperature ($T_\mathrm{bath}$) via $\frac{1}{2}P_\mathrm{J} = \mathcal{V}_\mathrm{S}\Sigma_\mathrm{e-ph}\left(T_\mathrm{e}^n-T_\mathrm{bath}^n\right),$ where $\mathcal{V}_\mathrm{S}$ is the volume of the S-island and $\Sigma_\mathrm{e-ph}$ is the e-ph coupling strength in Al ($n\approx5$). The factor $1/2$ takes into account that half of the Joule heat is dissipated in the N-terminals. Note that the above relation is approximate and neglects the effect  of Andreev reflection, which limits electronic heat conduction at near gap energies~\cite{Andreev}. This approximation works well for NW source-drain voltages sufficiently above the superconducting gap, as in the present experiment. It follows that the superconductivity in the S-island vanishes when a critical value $P_\mathrm{J}^\mathrm{c}$ is reached, which corresponds to the superconducting  critical temperature $T_\mathrm{e}=T_\mathrm{c}$. The Joule spectroscopy measures $P_\mathrm{J}^\mathrm{c}$ as a function of the magnetic field ($B$) directed along the NW and gives reasonable numbers~\cite{shpagina2024,ibabe2024} for $\Sigma_\mathrm{e-ph}$. Still, the $\Sigma_\mathrm{e-ph}$ obtained in this way in freely suspended devices~\cite{shpagina2024,shpagina2025} may fluctuate by $\pm$15\%, indicating a problem of the 2TM.    

A more general heat balance expression, referred to as the three-temperature model (3TM) below, takes into account that the phonon temperature in the S-island may deviate from the bath~\cite{roukes1985}, $T_\mathrm{ph}>T_\mathrm{bath}$, owing to a finite thermal conductance of the environment:
\begin{equation}
\frac{1}{2}P_\mathrm{J} = \mathcal{V}_\mathrm{S}\Sigma_\mathrm{e-ph}\left(T_\mathrm{e}^n-T_\mathrm{ph}^n\right) = \Sigma_{\mathrm{env}}\left(T_\mathrm{ph}^m-T_\mathrm{bath}^m \right) ,
\label{eq2}
\end{equation}
where $\Sigma_{\mathrm{env}}$ is the environmental  cooling parameter. Both $\Sigma_{\mathrm{env}}$ and the power law exponent $m$ depend on the details of thermal transport in the substrate or  surrounding liquid/gas. In the limit of $\Sigma_{\mathrm{env}}\rightarrow\infty$ one obtains $T_\mathrm{ph}=T_\mathrm{bath}$ as in the 2TM, otherwise the two temperatures are different and $T_\mathrm{ph}>T_\mathrm{bath}$, manifesting phonon heating owing to a thermal bottleneck effect.

In this work, we investigate the heat balance in S-islands of freely-suspended epitaxial full-shell Al/InAs NWs by means of the Joule spectroscopy as a function of $T_\mathrm{c}$ (tuned with the $B$-field) and $T_\mathrm{bath}$. We use resistive heaters in the vicinity of the NW device as a separate heat source, intended to vary the bath temperature, and calibrate the effect of the heater current ($I_\mathrm{H}$) using Johnson noise thermometry. The evolution of the  $P_\mathrm{J}^\mathrm{c}$ with $T_\mathrm{c}$ and $T_\mathrm{bath}$ contradicts the 2TM and manifests a sizable thermal bottleneck. We quantify $\Sigma_{\mathrm{env}}$, which is reasonably consistent with the thermal exchange with the surrounding $^3\mathrm{He}$.  

\begin{figure}[t!]
\begin{center}
\vspace{0mm}
  \includegraphics[width=1\linewidth]{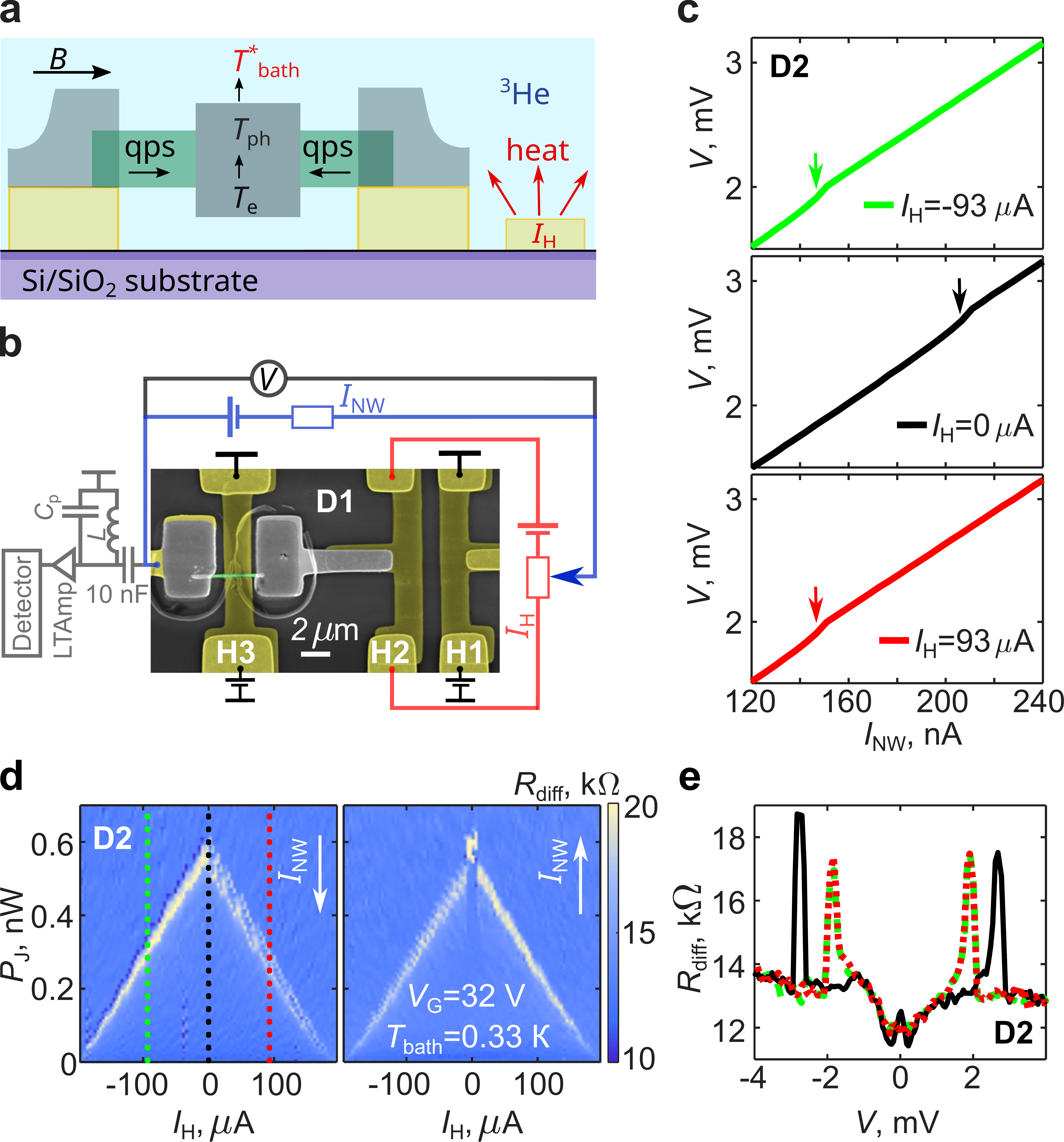}
\end{center}
  \caption{Device layout and impact of heaters. 
  	(a): Sketch of thermal pathways in a freely-suspended NW device. \add{The InAs core (green) is covered by an epitaxial Al shell (grey). The NW lays on the Au pads (yellow) and is freely suspended above the substrate (violet) and the middle heater H3 (yellow), which was not used in the present experiment. The surrounding $^3$He bath is shown in light blue.} Quasiparticles \add{(qps)} enter the superconductor  and heat the electron system (temperature $T_\text{e}$). The electrons cool to the phonon system in the S-island ($T_\text{ph}$)  and phonons cool to the bath. The effective bath temperature ($T_\mathrm{bath}^*$) is controlled with the nearby heater.
  	(b): False color scanning electron micrograph with the electrical circuit. H1 and H3 are non-contact heaters; H2 is a contact heater connected to one of the NW terminals. The other NW terminal is connected to a low-temperature amplifier (LTAmp) for noise measurement. 
  	(c): $I_\mathrm{NW}$-$V$ curves near the superconductor-normal transition in the device D2 for the three indicates values of $I_\mathrm{H}$. The transition points are marked with arrows. 
  	(d): Color-scale plot of the NW differential resistance  as a function of Joule power and heater current. The sweep directions of the NW transport current are indicated with arrows.  Dotted lines mark the $I_{\text{H}}$ values from (c). The back gate voltage and the bath temperature are shown in the legend.
  	(e): $R_{\text{diff}}$ versus $V$ for the data from panel (c) (same colors).} 
	\label{Fig1}
\end{figure}

Our devices are based on epitaxially grown Al/InAs NWs, with InAs core of $\sim150$\,nm diameter and Al shell thickness of $\approx40$\,nm \cite{shpagina2024, shpagina2025}. The NWs are placed on $\approx150$\,nm thick Au pads with a home-made micromanipulator and are freely suspended above the back-gated SiO$_2$/Si substrate. \add{This allows to minimize the thermal coupling of the S-island and the substrate, as compared to alternative approaches utilizing thin membranes~\cite{Baselmans2022} or thermally decoupling layers~\cite{Pimanov2025}.} A sketch of the device layout is given in Fig.~\ref{Fig1}a. The S-islands of the length 1-2.2\,$\mu$m  are formed by wet etching. In devices D2-D6 the ohmic contacts are formed via thermal evaporation of Cr/Au with ex situ passivation of the native oxide in ammonium polysulfide. Ohmic contacts in the device D1 were made with Al using in situ Ar milling. We observed no difference between devices with normal and superconducting contacts, which is not surprising in view of the bias voltage being much higher than the Al superconducting gap. A sketch of the electrical connections is depicted in Fig.~\ref{Fig1}b. Each of the two NW ohmic contacts is connected to a middle point of a $\sim 20$\,nm thick Au contact heater (H2 in Fig.~\ref{Fig1}b), with the heater resistance in the range $\sim1\mathrm{k}\Omega$.  Separate voltage sources  generate the NW transport current $I_\mathrm{NW}$ and the heater current $I_\mathrm{H}$. The contribution of the $I_\mathrm{H}$-dependent heater voltage to the measured NW source-drain voltage $V$ is accurately excluded by a compensation scheme~\cite{hoffmann2009}. In addition, we employ galvanically isolated non-contact heaters (H1 and H3 in Fig.~\ref{Fig1}b). Only one heater is used at a time. In the device D1, one of the NW ohmic contacts is connected to a tank circuit and a home-made low temperature radio-frequency amplifier for noise themometry measurements. The Joule heat is calculated as $P_\mathrm{J} = I_\mathrm{NW}V$. The $B$-field is nominally directed along the NW axis, with a possible misorientation angle of $\alpha<10^\circ$. The experiments are performed in a $^3\mathrm{He}$ cryostat with the device in $^3\mathrm{He}$ chamber. 

Fig.~\ref{Fig1}c details the $I_\mathrm{NW}$-$V$ behavior near the voltage switches at the superconductor-normal transition in device D2 ($B$=0, contact heater). The middle panel corresponds to $I_\mathrm{H}$=$0$ and for two other panels $I_\mathrm{H}$=$\pm93\,\mu$A. This data is obtained for down-sweeps from $I_\mathrm{NW}>0$ to zero. At increasing $|I_\mathrm{H}|$ the switch position shifts to smaller $I_\mathrm{NW}$, which corresponds to a decrease of the critical Joule power $P_\mathrm{J}^\mathrm{c}$. Fig.~\ref{Fig1}e shows the differential resistance $R_\mathrm{diff}\equiv dV/dI_\mathrm{NW}$ in dependence on the $I_\mathrm{NW}$ extracted via numerical differentiation. \add{The low bias behavior of the $R_\mathrm{diff}$ originates from the superconducting proximity effect in diffusive NS junctions, known as the reentrant resistance effect~\cite{Artemenko1979, Volkov1996, Nazarov1996, Lachenmann1997, Golubov1997, Kozhevnikov2000}, perhaps with a contribution of nonideal interfaces~\cite{bubis2017}.} \add{We are interested in the high bias features associated with voltage switches on the $I_\mathrm{NW}$-$V$ curves.} The sharp \add{maxima} of $R_\mathrm{diff}$ originating from \add{such} switches are indistinguishable for $I_\mathrm{H}$=$\pm93\,\mu$A (dashed and dotted lines) signaling a near perfect compensation of the heater circuit (see the SM). The full dependence of $P_\mathrm{J}^\mathrm{c}$ on the heater current is evident from the color-scale plots of Fig.~\ref{Fig1}d. Here we show $R_\mathrm{diff}$ as a function of the NW Joule power and $I_\mathrm{H}$ obtained both for $I_\mathrm{NW}>0$ (left panel) and for $I_\mathrm{NW}<0$ (right panel). The $I_\mathrm{NW}$ sweep directions are marked with vertical arrows. Bright yellow lines on the plot are formed by the $R_\mathrm{diff}$ maxima and demonstrate near linear decrease of $P_\mathrm{J}^\mathrm{c}$ with increasing $|I_\mathrm{H}|$. A slight deviation between the two datasets around $I_\mathrm{H}=0$ comes from the previously reported bistability effect~\cite{shpagina2025}, which quickly disappears at increasing $|I_\mathrm{H}|$. 

\begin{figure}[t!]
\begin{center}
\vspace{0mm}
  \includegraphics[width=1\linewidth]{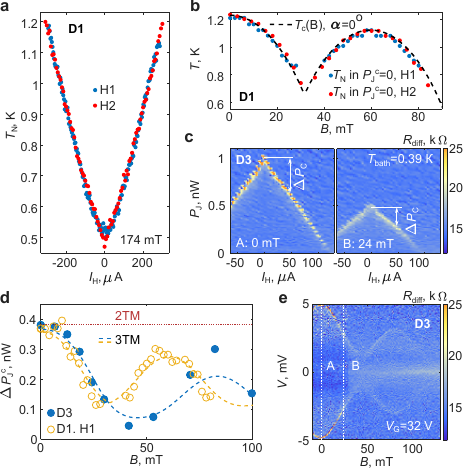}
\end{center}
  \caption{Joule spectroscopy with heaters.
  	(a): Johnson noise calibration: noise temperature $T_{\text{N}}$ vs.\ $I_{\text{H}}$ in the normal state ($174\,\text{mT}$) for the two heaters in the device D1.
  	(b): Symbols show $T_{\text{N}}$ corresponding to the $I_{\text{H}}$ needed to suppress the superconductivity in a given $B$-field ($P_\mathrm{J}^\mathrm{c}$=$0$). The dashed line shows the calculated $T_\mathrm{c}$ of the Al-shell.
  	(c): Color-scale plots of $R_{\text{diff}}$ vs.\ $P_{\text{J}}$ and $I_{\text{H}}$ for the device D3 in $B$=$0$ (left) and $B$=24\,mT (right). Scales show the reduction of the critical Joule power $\Delta P_{\text{J}}^{\text{c}}$ for $I_{\text{H}}\approx50\,\mu$A.
  	(d): $\Delta P_{\text{J}}^{\text{c}}$ vs.\ $B$ at fixed $I_\mathrm{H}$. $I_{\text{H}}\approx50\,\mu$A in D3 and $I_{\text{H}}\approx164\,\mu$A in D1. Dotted line: 2TM prediction; dashed lines: 3TM fits.
  	(e): Color-scale plot of $R_{\text{diff}}$ vs.\ $B$ and bias voltage $V$ in the device D3, showing Little--Parks oscillations. Dotted lines mark the $B$-field values from (c).} 
	\label{Fig2}
\end{figure}

To understand the thermal balance we calibrate the heaters' performance via noise thermometry, similar to previous experiments in InAs NWs~\cite{tikhonov2016a,tikhonov2016b}, see the SM for experimental details. The dependence of the NW noise temperature $T_\mathrm{N}(I_\mathrm{H})$ in device D1 is displayed in Fig.~\ref{Fig2}a. This data is obtained in a magnetic field $B=174\,$mT in the normal state of the S-island. For both heaters H1 and H2 a near linear increase of the $T_\mathrm{N}$ with $|I_\mathrm{H}|$ is observed. The similarity between the contact and non-contact heaters indicate that the heating of a freely-suspended device occurs via a surrounding $^3\mathrm{He}$. Note that our heaters rest on an amorphous $\mathrm{SiO}_2$ substrate and the observed linear trend $T_\mathrm{N}$ vs $|I_\mathrm{H}|$ is likely related to the pecularities of thermal relaxation in metal films via diffuson lattice excitations~\cite{baeva2021}. 

In Fig.~\ref{Fig2}b we compare the superconducting critical temperature evaluated from the Joule spectroscopy and noise calibration with the theoretical value for different $B$-field values. We extrapolate the dependence $P_\mathrm{J}^\mathrm{c}(I_\mathrm{H})$ to zero and find the heater current, which corresponds to the superconductivity suppressed in $I_\mathrm{NW}=0$. This value is converted to $T_\mathrm{N}$ by means of Fig.~\ref{Fig2}a and plotted as a function of $B$ for both heaters. The data closely reproduce the behavior of the $T_\mathrm{c}(B)$ calculated from the Al-shell parameters (dashed line). Fig.~\ref{Fig2}b therefore strongly suggests that the heat balance in our devices can be understood assuming that the heaters modify the effective bath temperature seen by the NW $T_\mathrm{bath}^*(I_\mathrm{H})> T_\mathrm{bath}$. In the following we analyze the data on this basis. 

Color-scale plots of Fig.~\ref{Fig2}c illustrate the change in the behavior of $P_\mathrm{J}^\mathrm{c}(I_\mathrm{H})$ with the $B$-field in device D3. The two panels correspond to $B$=$0$ (left) and  $B$=$24$\,mT (right), the values of the $B$-field within the first Little-Parks (LP) lobe, see the corresponding color-scale plot of $R_\mathrm{diff}$ in  Fig.~\ref{Fig2}e. Along with the overall decrease of $P_\mathrm{J}^\mathrm{c}$ caused by the decrease of the $T_\mathrm{c}(B)$, the slope of the dependence  $P_\mathrm{J}^\mathrm{c}(I_\mathrm{H})$  also diminishes. To demonstrate this, we analyze the change $\Delta P_\mathrm{J}^\mathrm{c}$ caused by a fixed value of the $I_\mathrm{H}\approx50\,\mu$A, see the scales in Fig.~\ref{Fig2}c. The full $\Delta P_\mathrm{J}^\mathrm{c}(B)$ dependence is shown in Fig.~\ref{Fig2}d (filled circles). $\Delta P_\mathrm{J}^\mathrm{c}$ oscillates with the $B$-field concurrently with the LP oscillations of the $T_\mathrm{c}(B)$. Similar behavior is observed in all devices and is shown with empty circles for the device D1 (H1). The oscillations of the $\Delta P_\mathrm{J}^\mathrm{c}(B)$ evidence the failure of the 2TM, which predicts $\Delta P_\mathrm{J}^\mathrm{c}(B)\propto (T_\mathrm{bath}^*)^n-T_\mathrm{bath}^n$ and independent of $B$ (dotted line). In the 3TM the $T_\mathrm{ph}$ at the transition point depends on the $B$-field via $T_\mathrm{c}(B)$, which allows to capture the experiment much better (dashed lines). Below we investigate the performance of 3TM in detail.
\begin{figure}[t]
\begin{center}
\vspace{0mm}
  \includegraphics[width=1\linewidth]{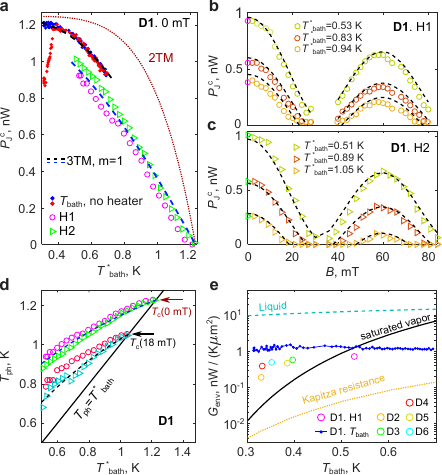}
\end{center}
  \caption{Thermal bottleneck effect.   	(a): $P_{\text{J}}^{\text{c}}$ vs.\ $T_{\text{bath}}^{*}$ in D1 in $B$=0. Open symbols show the data with heaters and solid symbols show the for data bath heating. Dotted line: 2TM; dashed: 3TM fits with $m$=1.
  	(b)-(c): $P_{\text{J}}^{\text{c}}$ vs.\ $B$ for heaters H1, H2 with the 3TM fits (D1).
  	(d): Dependence of the phonon temperature on $T_{\text{bath}}^{*}$ for the two indicated $B$-fields in the 3TM (D1). Symbols are extracted from the experiment using Eq.~(\ref{eq2}), the same style as in (b)-(c); dashed lines are the 3TM fits.
  	(e): Environmental thermal conductivity evaluated at $I_\mathrm{H}$=0 in dependence on $T_{\text{bath}}$. Data is compared with the  model fits (see legend). } 
	\label{Fig3}
\end{figure}

Fig.~\ref{Fig3}a displays the dependence $P_\mathrm{J}^\mathrm{c}(T_\mathrm{bath}^*)$ in the device D1 (open symbols), where we used Fig.~\ref{Fig2}a and $T_\mathrm{bath}^*(I_\mathrm{H})=T_\mathrm{N}(I_\mathrm{H})$. Almost linear decay of the $P_\mathrm{J}^\mathrm{c}$  with the $T_\mathrm{bath}^*$ is very far from the prediction of the 2TM (dotted line). By contrast, the 3TM captures the data correctly with the Eq.~(\ref{eq2}) with the exponent $m\approx 1-1.5$. We have chosen a slightly better fit with $m$=$1$ and $\Sigma_\mathrm{env}\approx1.2\,\mathrm{nW/K}$, see the dashed line. As shown in Figs.~\ref{Fig3}b and~\ref{Fig3}c, the dependence on the $B$-field is also consistent with the 3TM for both heaters  with the same $\Sigma_\mathrm{env}$. The $T_\mathrm{ph}$ extracted from the Eq.~(\ref{eq2}) is plotted as a function of $T_\mathrm{bath}^*$ in Fig.~\ref{Fig3}d for the two values of the $B$-field (symbols). Here, the thin solid line corresponds to $T_\mathrm{ph}$=$T_\mathrm{bath}^*$  in the 2TM. A sizeable thermal bottleneck effect is seen with $T_\mathrm{c}>T_\mathrm{ph}>T_\mathrm{bath}^*$, which is explained by the 3TM (dashed lines). 

The obtained exponent $m$=$1$ in the 3TM differs considerably from $m$=$4$ naively expected in a Kapitza resistance scenario~\cite{swartz1989}. We estimate that possible contribution of thermal conductivity via the InAs core, both the lattice and electronic, is negligible in our experiment (see the SM and  \footnote{A relatively weak thermal coupling of the S-island to the
NW contact pads as compared to the $^3$He bath makes any
possible asymmetry of the pad temperatures for $I_\mathrm{H}\neq 0$
unimportant.}). Thus, the thermal bottleneck effect is related to the physics of cooling a metallic island in $^3\mathrm{He}$. In order to discuss the possible scenarios, in Fig.~\ref{Fig3}e we summarize the obtained results in the form of environmental thermal conductivity evaluated at $P_\mathrm{J}=0$: $G_\mathrm{env}\equiv m\Sigma_\mathrm{env}T_\mathrm{bath}^{m-1}/A$, where $A$ is the surface area of the S-island. Open symbols correspond to experiments with heaters in different devices performed at slightly varying $T_\mathrm{bath}$. These variations were unintentional and depended on a prehistory of the $^3\mathrm{He}$ condensation and pumping in our setup. The $G_\mathrm{env}$ is roughly consistent among the different devices and weakly depends on $T_\mathrm{bath}$. The solid line is the prediction for cooling in $^3\mathrm{He}$ gas with $m$=$1$ and $G_\mathrm{env}=\frac{3}{8}\mathrm{P}_\mathrm{He}\sqrt{3k_\mathrm{B}/\left(m_\mathrm{He}T_\mathrm{bath}\right)}$, where $\mathrm{P}_\mathrm{He}$ is the saturated vapor pressure at a given $T_\mathrm{bath}$. In this model we took into account that the mean-free path of a $^3\mathrm{He}$ atom exceeds the dimensions of the S-island and assumed full thermalization of colliding atoms with the metal surface, see the SM. The dashed line is the prediction for cooling in liquid $^3\mathrm{He}$, for which $m$=$1.5$ and $G_\mathrm{env}\approx\kappa_\mathrm{L}(T_\mathrm{bath})/(r[\ln(L/2r)+1])$, where $L$ and $r$ are, respectively, the length and the outer radius of the S-island and $\kappa_\mathrm{L}(T)\propto T^{0.5}$ is the thermal conductivity of the liquid $^3\mathrm{He}$~\citep{peshkov1968,behnia2024}. Finally, the dotted line is the experimental data for the Kapitza resistance between copper and liquid $^3\mathrm{He}$ in our temperature range, taken from Ref.~\citep{anderson1972}. Obviously, none of the theoretical models is compatible with the experiment, although the functional dependence $G_\mathrm{env}(T_\mathrm{bath})$ is closer to the cooling via liquid.
\begin{figure}[t]
\begin{center}
\vspace{0mm}
  \includegraphics[width=1\linewidth]{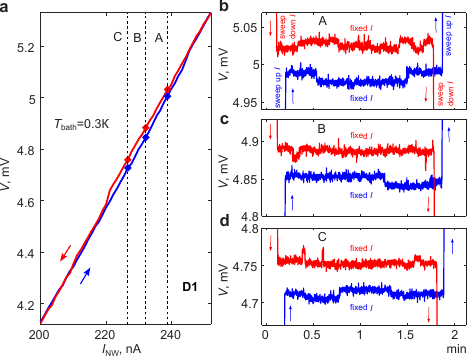}
\end{center}
  \caption{Time-resolved $I_{\text{NW}}$--$V$ measurement.
  	(a): Hysteretic loop in D1 at $T_{\text{bath}}=0.3\,\text{K}$. Blue/red solid lines correspond to up/down sweeps of $I_{\text{NW}}$ ($\sim 3$\,s per sweep).
  	(b)-(d): Time dependence of the voltage signal when paused at the points A, B, and C inside the loop, see the dots in (a). } 
	\label{Fig4}
\end{figure}

We also performed experiments by changing the temperature of the bath as a whole, as shown by solid symbols in Fig.~\ref{Fig3}a. Different symbols correspond to different $I_\mathrm{NW}$ sweep directions in the device D1. At the lowest $T_\mathrm{bath}\approx0.3$\,K a previously reported giant hysteresis is found~\cite{shpagina2025}, associated with a quasiparticle non-equilibrium in the S-island~\cite{keizer2006,snyman2009,kawamura2024,kawamura2025a,kawamura2026}. At increasing $T_\mathrm{bath}$, which we could trace up to 0.7\,K, the hysteresis quickly vanishes and the $P_\mathrm{J}^\mathrm{c}$ diminishes. The higher $P_\mathrm{J}^\mathrm{c}$, as compared to the experiment with the heaters, corresponds to a better environmental cooling in the 3TM ($m$=1, thick dashed line). Corresponding dependence $G_\mathrm{env}(T_\mathrm{bath})$ is shown in Fig.~\ref{Fig3}e by solid symbols. Overall, our observations are consistent with the cooling of the S-island via a thin film of liquid $^3\mathrm{He}$. In this scenario, both the data scatter in various experiments and the reduced $G_\mathrm{env}$ as compared to the prediction for bulk liquid $^3\mathrm{He}$ in Fig.~\ref{Fig3}e are explainable by variations of the thickness of a wetting layer. This hypothesis is consistent with the estimates of the level of liquid $^3\mathrm{He}$, which is usually a few millimeters below the device. Other indications, including uneven behaviors of the 
$P_\mathrm{J}^\mathrm{c}(I_\mathrm{H})$ sometimes observed in experiments with the heaters are summarized in the SM.  

The sensitivity of the $P_\mathrm{J}^\mathrm{c}$ to the change of $\Sigma_\mathrm{env}$ poses a question as to whether the giant hysteresis at the lowest $T_\mathrm{bath}$ may be an artifact of boiling in the vicinity of the S-island. We rule out this possibility by time-resolved $I$-$V$ measurements. Fig.~\ref{Fig4}a displays the hysteresis loop in the device D1 measured at $T_\mathrm{bath}$=$0.3$\,K. The blue/red solid lines correspond to the up/down sweeps of the $I_\mathrm{NW}$ obtained in about 3 seconds per indicated sweep range. Figs.~\ref{Fig4}b, ~\ref{Fig4}c and ~\ref{Fig4}d show the data obtained with intermediate stops inside the hysteresis loop at the three currents indicated by vertical dashed lines A, B and C in Fig.~\ref{Fig4}a, respectively. During each stop, the $I_\mathrm{NW}$ was fixed for $\sim 1.5$~minutes and no switching between the up-sweep and down-sweep branches occurred. The difference in voltages measured for the two sweep directions remained close to the data of the coarse sweep, see symbols in Fig.~\ref{Fig4}a, up to a contribution of the random telegraph noise in the normal parts of the InAs NW, see Figs.~\ref{Fig4}b-\ref{Fig4}d. This demonstrates that the observed hysteresis is a stationary non-equilibrium phenomenon not related to any transient  effects in the surrounding $^3\mathrm{He}$.   

A direct consequence of our observations is that the 2TM used in our previous work~\cite{shpagina2024,shpagina2025} underestimated the value of the electron-phonon cooling power $\Sigma_\mathrm{e-ph}$. The possible mistake is within 20\% because of the large value of the exponent $n=5$ in the Eq.~(\ref{eq2}), which makes the contribution of the finite  $T_\mathrm{ph}$ small unless it is very close to the $T_\mathrm{c}$. All the fits in the present work use the same value $\Sigma_\mathrm{e-ph}=5.2\,\mathrm{nW\mu m^{-3}K^{-5}}$.

In summary, we investigated the non-equilibrium superconductor-normal transition in freely suspended S-islands in epitaxial Al/InAs NWs by means of a Joule spectroscopy. The behavior of the critical Joule power of the transition with the heater current, the bath temperature and the $B$-field is inconsistent with the two-temperature relaxation model, manifesting sizable thermal bottleneck. The three-temperature model describes the data quantitatively and suggests that the S-island cools via heat conduction into the surrounding $^3\mathrm{He}$ liquid film. Our experiment shows potential of using the Joule spectroscopy approach for understanding of thermal relaxation at the nanoscale.  

We acknowledge a valuable discussion with E.M.~Baeva. This work
was supported by the Russian Science Foundation Project No.
26-12-00085.


\begin{thebibliography}{5}%
\makeatletter
\providecommand \@ifxundefined [1]{%
 \@ifx{#1\undefined}
}%
\providecommand \@ifnum [1]{%
 \ifnum #1\expandafter \@firstoftwo
 \else \expandafter \@secondoftwo
 \fi
}%
\providecommand \@ifx [1]{%
 \ifx #1\expandafter \@firstoftwo
 \else \expandafter \@secondoftwo
 \fi
}%
\providecommand \natexlab [1]{#1}%
\providecommand \enquote  [1]{``#1''}%
\providecommand \bibnamefont  [1]{#1}%
\providecommand \bibfnamefont [1]{#1}%
\providecommand \citenamefont [1]{#1}%
\providecommand \href@noop [0]{\@secondoftwo}%
\providecommand \href [0]{\begingroup \@sanitize@url \@href}%
\providecommand \@href[1]{\@@startlink{#1}\@@href}%
\providecommand \@@href[1]{\endgroup#1\@@endlink}%
\providecommand \@sanitize@url [0]{\catcode `\\12\catcode `\$12\catcode
  `\&12\catcode `\#12\catcode `\^12\catcode `\_12\catcode `\%12\relax}%
\providecommand \@@startlink[1]{}%
\providecommand \@@endlink[0]{}%
\providecommand \url  [0]{\begingroup\@sanitize@url \@url }%
\providecommand \@url [1]{\endgroup\@href {#1}{\urlprefix }}%
\providecommand \urlprefix  [0]{URL }%
\providecommand \Eprint [0]{\href }%
\providecommand \doibase [0]{https://doi.org/}%
\providecommand \selectlanguage [0]{\@gobble}%
\providecommand \bibinfo  [0]{\@secondoftwo}%
\providecommand \bibfield  [0]{\@secondoftwo}%
\providecommand \translation [1]{[#1]}%
\providecommand \BibitemOpen [0]{}%
\providecommand \bibitemStop [0]{}%
\providecommand \bibitemNoStop [0]{.\EOS\space}%
\providecommand \EOS [0]{\spacefactor3000\relax}%
\providecommand \BibitemShut  [1]{\csname bibitem#1\endcsname}%
\let\auto@bib@innerbib\@empty
\bibitem [{\citenamefont {Shpagina}\ \emph {et~al.}(2024)\citenamefont
  {Shpagina}, \citenamefont {Tikhonov}, \citenamefont {Ruhstorfer},
  \citenamefont {Koblm{\"u}ller},\ and\ \citenamefont
  {Khrapai}}]{shpagina2024}%
  \BibitemOpen
  \bibfield  {author} {\bibinfo {author} {\bibfnamefont {E.~V.}\ \bibnamefont
  {Shpagina}}, \bibinfo {author} {\bibfnamefont {E.~S.}\ \bibnamefont
  {Tikhonov}}, \bibinfo {author} {\bibfnamefont {D.}~\bibnamefont
  {Ruhstorfer}}, \bibinfo {author} {\bibfnamefont {G.}~\bibnamefont
  {Koblm{\"u}ller}},\ and\ \bibinfo {author} {\bibfnamefont {V.~S.}\
  \bibnamefont {Khrapai}},\ }\bibfield  {title} {\bibinfo {title} {{\flqq Fate
  of the Superconducting State in Floating Islands of Hybrid Nanowire
  Devices\frqq}},\ }\href {https://doi.org/10.1103/PhysRevB.109.L140501}
  {\bibfield  {journal} {\bibinfo  {journal} {Physical Review B}\ }\textbf
  {\bibinfo {volume} {109}},\ \bibinfo {pages} {L140501} (\bibinfo {year}
  {2024})}\BibitemShut {NoStop}%
\bibitem [{\citenamefont {Shpagina}\ \emph {et~al.}(2025)\citenamefont
  {Shpagina}, \citenamefont {Tikhonov}, \citenamefont {Ruhstorfer},
  \citenamefont {Koblm\"uller},\ and\ \citenamefont {Khrapai}}]{shpagina2025}%
  \BibitemOpen
  \bibfield  {author} {\bibinfo {author} {\bibfnamefont {E.~V.}\ \bibnamefont
  {Shpagina}}, \bibinfo {author} {\bibfnamefont {E.~S.}\ \bibnamefont
  {Tikhonov}}, \bibinfo {author} {\bibfnamefont {D.}~\bibnamefont
  {Ruhstorfer}}, \bibinfo {author} {\bibfnamefont {G.}~\bibnamefont
  {Koblm\"uller}},\ and\ \bibinfo {author} {\bibfnamefont {V.~S.}\ \bibnamefont
  {Khrapai}},\ }\bibfield  {title} {\bibinfo {title} {{\flqq Superconducting
  bistability in floating Al islands of hybrid Al/InAs nanowires\frqq}},\
  }\href {https://doi.org/10.1103/l4tq-bxq9} {\bibfield  {journal} {\bibinfo
  {journal} {Phys. Rev. B}\ }\textbf {\bibinfo {volume} {112}},\ \bibinfo
  {pages} {L140503} (\bibinfo {year} {2025})}\BibitemShut {NoStop}%
\bibitem [{\citenamefont {{E S Tikhonov}}\ \emph {et~al.}(2016)\citenamefont
  {{E S Tikhonov}}, \citenamefont {{D V Shovkun}}, \citenamefont {{D
  Ercolani}}, \citenamefont {{F Rossella}}, \citenamefont {{M Rocci}},
  \citenamefont {{L Sorba}}, \citenamefont {{S Roddaro}},\ and\ \citenamefont
  {{V S Khrapai}}}]{Tikhonov2016a}%
  \BibitemOpen
  \bibfield  {author} {\bibinfo {author} {\bibnamefont {{E S Tikhonov}}},
  \bibinfo {author} {\bibnamefont {{D V Shovkun}}}, \bibinfo {author}
  {\bibnamefont {{D Ercolani}}}, \bibinfo {author} {\bibnamefont {{F
  Rossella}}}, \bibinfo {author} {\bibnamefont {{M Rocci}}}, \bibinfo {author}
  {\bibnamefont {{L Sorba}}}, \bibinfo {author} {\bibnamefont {{S Roddaro}}},\
  and\ \bibinfo {author} {\bibnamefont {{V S Khrapai}}},\ }\bibfield  {title}
  {\bibinfo {title} {{\flqq Local Noise in a Diffusive Conductor\frqq}},\
  }\href {https://doi.org/10.1038/srep30621} {\bibfield  {journal} {\bibinfo
  {journal} {Scientific Reports}\ }\textbf {\bibinfo {volume} {6}},\ \bibinfo
  {pages} {30621} (\bibinfo {year} {2016})}\BibitemShut {NoStop}%
\bibitem [{\citenamefont {Tikhonov}(2016)}]{Tikhonov2016b}%
  \BibitemOpen
  \bibfield  {author} {\bibinfo {author} {\bibfnamefont {E.~S.}\ \bibnamefont
  {Tikhonov}},\ }\bibfield  {title} {\bibinfo {title} {{\flqq Noise Thermometry
  Applied to Thermoelectric Measurements in {{InAs}} Nanowires\frqq}},\ }\href
  {https://doi.org/10.1088/0268-1242/31/10/104001} {\bibfield  {journal}
  {\bibinfo  {journal} {Semicond. Sci. Technol.}\ }\textbf {\bibinfo {volume}
  {31}},\ \bibinfo {pages} {104001} (\bibinfo {year} {2016})}\BibitemShut
  {NoStop}%
\bibitem [{\citenamefont {Sydoriak}\ and\ \citenamefont
  {Roberts}(1957)}]{Sydoriak1957}%
  \BibitemOpen
  \bibfield  {author} {\bibinfo {author} {\bibfnamefont {S.~G.}\ \bibnamefont
  {Sydoriak}}\ and\ \bibinfo {author} {\bibfnamefont {T.~R.}\ \bibnamefont
  {Roberts}},\ }\bibfield  {title} {\bibinfo {title} {{\flqq Thermodynamic
  Properties of Liquid Helium Three. Vapor Pressures below 1°K\frqq}},\ }\href
  {https://doi.org/10.1103/PhysRev.106.175} {\bibfield  {journal} {\bibinfo
  {journal} {Physical Review}\ }\textbf {\bibinfo {volume} {106}},\ \bibinfo
  {pages} {175} (\bibinfo {year} {1957})}\BibitemShut {NoStop}%
\end{thebibliography}%


\begin{thebibliography}{41}%
\makeatletter
\providecommand \@ifxundefined [1]{%
 \@ifx{#1\undefined}
}%
\providecommand \@ifnum [1]{%
 \ifnum #1\expandafter \@firstoftwo
 \else \expandafter \@secondoftwo
 \fi
}%
\providecommand \@ifx [1]{%
 \ifx #1\expandafter \@firstoftwo
 \else \expandafter \@secondoftwo
 \fi
}%
\providecommand \natexlab [1]{#1}%
\providecommand \enquote  [1]{``#1''}%
\providecommand \bibnamefont  [1]{#1}%
\providecommand \bibfnamefont [1]{#1}%
\providecommand \citenamefont [1]{#1}%
\providecommand \href@noop [0]{\@secondoftwo}%
\providecommand \href [0]{\begingroup \@sanitize@url \@href}%
\providecommand \@href[1]{\@@startlink{#1}\@@href}%
\providecommand \@@href[1]{\endgroup#1\@@endlink}%
\providecommand \@sanitize@url [0]{\catcode `\\12\catcode `\$12\catcode
  `\&12\catcode `\#12\catcode `\^12\catcode `\_12\catcode `\%12\relax}%
\providecommand \@@startlink[1]{}%
\providecommand \@@endlink[0]{}%
\providecommand \url  [0]{\begingroup\@sanitize@url \@url }%
\providecommand \@url [1]{\endgroup\@href {#1}{\urlprefix }}%
\providecommand \urlprefix  [0]{URL }%
\providecommand \Eprint [0]{\href }%
\providecommand \doibase [0]{https://doi.org/}%
\providecommand \selectlanguage [0]{\@gobble}%
\providecommand \bibinfo  [0]{\@secondoftwo}%
\providecommand \bibfield  [0]{\@secondoftwo}%
\providecommand \translation [1]{[#1]}%
\providecommand \BibitemOpen [0]{}%
\providecommand \bibitemStop [0]{}%
\providecommand \bibitemNoStop [0]{.\EOS\space}%
\providecommand \EOS [0]{\spacefactor3000\relax}%
\providecommand \BibitemShut  [1]{\csname bibitem#1\endcsname}%
\let\auto@bib@innerbib\@empty
\bibitem [{\citenamefont {Akhmerov}\ \emph {et~al.}(2011)\citenamefont
  {Akhmerov}, \citenamefont {Dahlhaus}, \citenamefont {Hassler}, \citenamefont
  {Wimmer},\ and\ \citenamefont {Beenakker}}]{akhmerov2011}%
  \BibitemOpen
  \bibfield  {author} {\bibinfo {author} {\bibfnamefont {A.~R.}\ \bibnamefont
  {Akhmerov}}, \bibinfo {author} {\bibfnamefont {J.~P.}\ \bibnamefont
  {Dahlhaus}}, \bibinfo {author} {\bibfnamefont {F.}~\bibnamefont {Hassler}},
  \bibinfo {author} {\bibfnamefont {M.}~\bibnamefont {Wimmer}},\ and\ \bibinfo
  {author} {\bibfnamefont {C.~W.~J.}\ \bibnamefont {Beenakker}},\ }\bibfield
  {title} {\bibinfo {title} {Quantized {{Conductance}} at the {{Majorana Phase
  Transition}} in a {{Disordered Superconducting Wire}}},\ }\href
  {https://doi.org/10.1103/PhysRevLett.106.057001} {\bibfield  {journal}
  {\bibinfo  {journal} {Physical Review Letters}\ }\textbf {\bibinfo {volume}
  {106}},\ \bibinfo {pages} {057001} (\bibinfo {year} {2011})}\BibitemShut
  {NoStop}%
\bibitem [{\citenamefont {Pan}\ \emph {et~al.}(2021)\citenamefont {Pan},
  \citenamefont {Sau},\ and\ \citenamefont {Das~Sarma}}]{pan2021}%
  \BibitemOpen
  \bibfield  {author} {\bibinfo {author} {\bibfnamefont {H.}~\bibnamefont
  {Pan}}, \bibinfo {author} {\bibfnamefont {J.~D.}\ \bibnamefont {Sau}},\ and\
  \bibinfo {author} {\bibfnamefont {S.}~\bibnamefont {Das~Sarma}},\ }\bibfield
  {title} {\bibinfo {title} {Three-terminal nonlocal conductance in
  {{Majorana}} nanowires: {{Distinguishing}} topological and trivial in
  realistic systems with disorder and inhomogeneous potential},\ }\href
  {https://doi.org/10.1103/PhysRevB.103.014513} {\bibfield  {journal} {\bibinfo
   {journal} {Physical Review B}\ }\textbf {\bibinfo {volume} {103}},\ \bibinfo
  {pages} {014513} (\bibinfo {year} {2021})}\BibitemShut {NoStop}%
\bibitem [{\citenamefont {Bubis}\ \emph {et~al.}(2021)\citenamefont {Bubis},
  \citenamefont {Shpagina}, \citenamefont {Nasibulin},\ and\ \citenamefont
  {Khrapai}}]{bubis2021}%
  \BibitemOpen
  \bibfield  {author} {\bibinfo {author} {\bibfnamefont {A.~V.}\ \bibnamefont
  {Bubis}}, \bibinfo {author} {\bibfnamefont {E.~V.}\ \bibnamefont {Shpagina}},
  \bibinfo {author} {\bibfnamefont {A.~G.}\ \bibnamefont {Nasibulin}},\ and\
  \bibinfo {author} {\bibfnamefont {V.~S.}\ \bibnamefont {Khrapai}},\
  }\bibfield  {title} {\bibinfo {title} {Thermal conductance and nonequilibrium
  superconductivity in a diffusive {{NSN}} wire probed by shot noise},\ }\href
  {https://doi.org/10.1103/PhysRevB.104.125409} {\bibfield  {journal} {\bibinfo
   {journal} {Physical Review B}\ }\textbf {\bibinfo {volume} {104}},\ \bibinfo
  {pages} {125409} (\bibinfo {year} {2021})}\BibitemShut {NoStop}%
\bibitem [{\citenamefont {Denisov}\ \emph {et~al.}(2021)\citenamefont
  {Denisov}, \citenamefont {Bubis}, \citenamefont {Piatrusha}, \citenamefont
  {Titova}, \citenamefont {Nasibulin}, \citenamefont {Becker}, \citenamefont
  {Treu}, \citenamefont {Ruhstorfer}, \citenamefont {Koblm{\"u}ller},
  \citenamefont {Tikhonov},\ and\ \citenamefont {Khrapai}}]{denisov2021}%
  \BibitemOpen
  \bibfield  {author} {\bibinfo {author} {\bibfnamefont {A.~O.}\ \bibnamefont
  {Denisov}}, \bibinfo {author} {\bibfnamefont {A.~V.}\ \bibnamefont {Bubis}},
  \bibinfo {author} {\bibfnamefont {S.~U.}\ \bibnamefont {Piatrusha}}, \bibinfo
  {author} {\bibfnamefont {N.~A.}\ \bibnamefont {Titova}}, \bibinfo {author}
  {\bibfnamefont {A.~G.}\ \bibnamefont {Nasibulin}}, \bibinfo {author}
  {\bibfnamefont {J.}~\bibnamefont {Becker}}, \bibinfo {author} {\bibfnamefont
  {J.}~\bibnamefont {Treu}}, \bibinfo {author} {\bibfnamefont {D.}~\bibnamefont
  {Ruhstorfer}}, \bibinfo {author} {\bibfnamefont {G.}~\bibnamefont
  {Koblm{\"u}ller}}, \bibinfo {author} {\bibfnamefont {E.~S.}\ \bibnamefont
  {Tikhonov}},\ and\ \bibinfo {author} {\bibfnamefont {V.~S.}\ \bibnamefont
  {Khrapai}},\ }\bibfield  {title} {\bibinfo {title} {Charge-neutral nonlocal
  response in superconductor-{{InAs}} nanowire hybrid devices},\ }\href
  {https://doi.org/10.1088/1361-6641/ac187b} {\bibfield  {journal} {\bibinfo
  {journal} {Semicond. Sci. Technol.}\ }\textbf {\bibinfo {volume} {36}},\
  \bibinfo {pages} {09LT04} (\bibinfo {year} {2021})}\BibitemShut {NoStop}%
\bibitem [{\citenamefont {Denisov}\ \emph {et~al.}(2022)\citenamefont
  {Denisov}, \citenamefont {Bubis}, \citenamefont {Piatrusha}, \citenamefont
  {Titova}, \citenamefont {Nasibulin}, \citenamefont {Becker}, \citenamefont
  {Treu}, \citenamefont {Ruhstorfer}, \citenamefont {Koblm{\"u}ller},
  \citenamefont {Tikhonov},\ and\ \citenamefont {Khrapai}}]{denisov2022}%
  \BibitemOpen
  \bibfield  {author} {\bibinfo {author} {\bibfnamefont {A.}~\bibnamefont
  {Denisov}}, \bibinfo {author} {\bibfnamefont {A.}~\bibnamefont {Bubis}},
  \bibinfo {author} {\bibfnamefont {S.}~\bibnamefont {Piatrusha}}, \bibinfo
  {author} {\bibfnamefont {N.}~\bibnamefont {Titova}}, \bibinfo {author}
  {\bibfnamefont {A.}~\bibnamefont {Nasibulin}}, \bibinfo {author}
  {\bibfnamefont {J.}~\bibnamefont {Becker}}, \bibinfo {author} {\bibfnamefont
  {J.}~\bibnamefont {Treu}}, \bibinfo {author} {\bibfnamefont {D.}~\bibnamefont
  {Ruhstorfer}}, \bibinfo {author} {\bibfnamefont {G.}~\bibnamefont
  {Koblm{\"u}ller}}, \bibinfo {author} {\bibfnamefont {E.}~\bibnamefont
  {Tikhonov}},\ and\ \bibinfo {author} {\bibfnamefont {V.}~\bibnamefont
  {Khrapai}},\ }\bibfield  {title} {\bibinfo {title} {Heat-{{Mode Excitation}}
  in a {{Proximity Superconductor}}},\ }\href
  {https://doi.org/10.3390/nano12091461} {\bibfield  {journal} {\bibinfo
  {journal} {Nanomaterials}\ }\textbf {\bibinfo {volume} {12}},\ \bibinfo
  {pages} {1461} (\bibinfo {year} {2022})}\BibitemShut {NoStop}%
\bibitem [{\citenamefont {Majidi}\ \emph {et~al.}(2022)\citenamefont {Majidi},
  \citenamefont {Josefsson}, \citenamefont {Kumar}, \citenamefont {Leijnse},
  \citenamefont {Samuelson}, \citenamefont {Courtois}, \citenamefont
  {Winkelmann},\ and\ \citenamefont {Maisi}}]{majidi2022}%
  \BibitemOpen
  \bibfield  {author} {\bibinfo {author} {\bibfnamefont {D.}~\bibnamefont
  {Majidi}}, \bibinfo {author} {\bibfnamefont {M.}~\bibnamefont {Josefsson}},
  \bibinfo {author} {\bibfnamefont {M.}~\bibnamefont {Kumar}}, \bibinfo
  {author} {\bibfnamefont {M.}~\bibnamefont {Leijnse}}, \bibinfo {author}
  {\bibfnamefont {L.}~\bibnamefont {Samuelson}}, \bibinfo {author}
  {\bibfnamefont {H.}~\bibnamefont {Courtois}}, \bibinfo {author}
  {\bibfnamefont {C.~B.}\ \bibnamefont {Winkelmann}},\ and\ \bibinfo {author}
  {\bibfnamefont {V.~F.}\ \bibnamefont {Maisi}},\ }\bibfield  {title} {\bibinfo
  {title} {Quantum confinement suppressing electronic heat flow below the
  {{Wiedemann-Franz}} law},\ }\href
  {https://doi.org/10.1021/acs.nanolett.1c03437} {\bibfield  {journal}
  {\bibinfo  {journal} {Nano Letters}\ }\textbf {\bibinfo {volume} {22}},\
  \bibinfo {pages} {630} (\bibinfo {year} {2022})}\BibitemShut {NoStop}%
\bibitem [{\citenamefont {Haldar}\ \emph {et~al.}(2026)\citenamefont {Haldar},
  \citenamefont {Subero}, \citenamefont {Kumar}, \citenamefont {Karimi},
  \citenamefont {Burke}, \citenamefont {Samuelson}, \citenamefont {Pekola},\
  and\ \citenamefont {Maisi}}]{haldar2026}%
  \BibitemOpen
  \bibfield  {author} {\bibinfo {author} {\bibfnamefont {S.}~\bibnamefont
  {Haldar}}, \bibinfo {author} {\bibfnamefont {D.}~\bibnamefont {Subero}},
  \bibinfo {author} {\bibfnamefont {M.}~\bibnamefont {Kumar}}, \bibinfo
  {author} {\bibfnamefont {B.}~\bibnamefont {Karimi}}, \bibinfo {author}
  {\bibfnamefont {A.}~\bibnamefont {Burke}}, \bibinfo {author} {\bibfnamefont
  {L.}~\bibnamefont {Samuelson}}, \bibinfo {author} {\bibfnamefont
  {J.}~\bibnamefont {Pekola}},\ and\ \bibinfo {author} {\bibfnamefont {V.~F.}\
  \bibnamefont {Maisi}},\ }\href {https://arxiv.org/abs/2603.29358} {\bibinfo
  {title} {Heat conduction and energy relaxation in an {{InAs}} nanowire
  approaching the clean one-dimensional limit}} (\bibinfo {year} {2026}),\
  \Eprint {https://arxiv.org/abs/2603.29358} {arXiv:2603.29358
  [cond-mat.mes-hall]} \BibitemShut {NoStop}%
\bibitem [{\citenamefont {Ibabe}\ \emph {et~al.}(2023)\citenamefont {Ibabe},
  \citenamefont {G{\'o}mez}, \citenamefont {Steffensen}, \citenamefont {Kanne},
  \citenamefont {Nyg{\aa}rd}, \citenamefont {Yeyati},\ and\ \citenamefont
  {Lee}}]{ibabe2023}%
  \BibitemOpen
  \bibfield  {author} {\bibinfo {author} {\bibfnamefont {A.}~\bibnamefont
  {Ibabe}}, \bibinfo {author} {\bibfnamefont {M.}~\bibnamefont {G{\'o}mez}},
  \bibinfo {author} {\bibfnamefont {G.~O.}\ \bibnamefont {Steffensen}},
  \bibinfo {author} {\bibfnamefont {T.}~\bibnamefont {Kanne}}, \bibinfo
  {author} {\bibfnamefont {J.}~\bibnamefont {Nyg{\aa}rd}}, \bibinfo {author}
  {\bibfnamefont {A.~L.}\ \bibnamefont {Yeyati}},\ and\ \bibinfo {author}
  {\bibfnamefont {E.~J.~H.}\ \bibnamefont {Lee}},\ }\bibfield  {title}
  {\bibinfo {title} {Joule spectroscopy of hybrid superconductor--semiconductor
  nanodevices},\ }\href {https://doi.org/10.1038/s41467-023-38533-2} {\bibfield
   {journal} {\bibinfo  {journal} {Nature Communications}\ }\textbf {\bibinfo
  {volume} {14}},\ \bibinfo {pages} {2873} (\bibinfo {year}
  {2023})}\BibitemShut {NoStop}%
\bibitem [{\citenamefont {Liu}\ \emph {et~al.}(2023)\citenamefont {Liu},
  \citenamefont {Pan}, \citenamefont {Le}, \citenamefont {He}, \citenamefont
  {Jia}, \citenamefont {Zhu}, \citenamefont {Yang}, \citenamefont {Lyu},
  \citenamefont {Liu}, \citenamefont {Shen}, \citenamefont {Zhao},
  \citenamefont {Lu},\ and\ \citenamefont {Qu}}]{liu2023}%
  \BibitemOpen
  \bibfield  {author} {\bibinfo {author} {\bibfnamefont {M.-L.}\ \bibnamefont
  {Liu}}, \bibinfo {author} {\bibfnamefont {D.}~\bibnamefont {Pan}}, \bibinfo
  {author} {\bibfnamefont {T.}~\bibnamefont {Le}}, \bibinfo {author}
  {\bibfnamefont {J.-B.}\ \bibnamefont {He}}, \bibinfo {author} {\bibfnamefont
  {Z.-M.}\ \bibnamefont {Jia}}, \bibinfo {author} {\bibfnamefont
  {S.}~\bibnamefont {Zhu}}, \bibinfo {author} {\bibfnamefont {G.}~\bibnamefont
  {Yang}}, \bibinfo {author} {\bibfnamefont {Z.-Z.}\ \bibnamefont {Lyu}},
  \bibinfo {author} {\bibfnamefont {G.-T.}\ \bibnamefont {Liu}}, \bibinfo
  {author} {\bibfnamefont {J.}~\bibnamefont {Shen}}, \bibinfo {author}
  {\bibfnamefont {J.-H.}\ \bibnamefont {Zhao}}, \bibinfo {author}
  {\bibfnamefont {L.}~\bibnamefont {Lu}},\ and\ \bibinfo {author}
  {\bibfnamefont {F.-M.}\ \bibnamefont {Qu}},\ }\bibfield  {title} {\bibinfo
  {title} {Gate-{{Tunable Negative Differential Conductance}} in {{Hybrid
  Semiconductor}}--{{Superconductor Devices}}},\ }\href
  {https://doi.org/10.1088/0256-307X/40/6/067301} {\bibfield  {journal}
  {\bibinfo  {journal} {Chinese Physics Letters}\ }\textbf {\bibinfo {volume}
  {40}},\ \bibinfo {pages} {067301} (\bibinfo {year} {2023})}\BibitemShut
  {NoStop}%
\bibitem [{\citenamefont {Shpagina}\ \emph {et~al.}(2024)\citenamefont
  {Shpagina}, \citenamefont {Tikhonov}, \citenamefont {Ruhstorfer},
  \citenamefont {Koblm{\"u}ller},\ and\ \citenamefont
  {Khrapai}}]{shpagina2024}%
  \BibitemOpen
  \bibfield  {author} {\bibinfo {author} {\bibfnamefont {E.~V.}\ \bibnamefont
  {Shpagina}}, \bibinfo {author} {\bibfnamefont {E.~S.}\ \bibnamefont
  {Tikhonov}}, \bibinfo {author} {\bibfnamefont {D.}~\bibnamefont
  {Ruhstorfer}}, \bibinfo {author} {\bibfnamefont {G.}~\bibnamefont
  {Koblm{\"u}ller}},\ and\ \bibinfo {author} {\bibfnamefont {V.~S.}\
  \bibnamefont {Khrapai}},\ }\bibfield  {title} {\bibinfo {title} {Fate of the
  superconducting state in floating islands of hybrid nanowire devices},\
  }\href {https://doi.org/10.1103/PhysRevB.109.L140501} {\bibfield  {journal}
  {\bibinfo  {journal} {Physical Review B}\ }\textbf {\bibinfo {volume}
  {109}},\ \bibinfo {pages} {L140501} (\bibinfo {year} {2024})}\BibitemShut
  {NoStop}%
\bibitem [{\citenamefont {Ibabe}\ \emph {et~al.}(2024)\citenamefont {Ibabe},
  \citenamefont {Steffensen}, \citenamefont {Casal}, \citenamefont {G{\'o}mez},
  \citenamefont {Kanne}, \citenamefont {Nyg{\aa}rd}, \citenamefont
  {Levy~Yeyati},\ and\ \citenamefont {Lee}}]{ibabe2024}%
  \BibitemOpen
  \bibfield  {author} {\bibinfo {author} {\bibfnamefont {{\'A}.}~\bibnamefont
  {Ibabe}}, \bibinfo {author} {\bibfnamefont {G.~O.}\ \bibnamefont
  {Steffensen}}, \bibinfo {author} {\bibfnamefont {I.}~\bibnamefont {Casal}},
  \bibinfo {author} {\bibfnamefont {M.}~\bibnamefont {G{\'o}mez}}, \bibinfo
  {author} {\bibfnamefont {T.}~\bibnamefont {Kanne}}, \bibinfo {author}
  {\bibfnamefont {J.}~\bibnamefont {Nyg{\aa}rd}}, \bibinfo {author}
  {\bibfnamefont {A.}~\bibnamefont {Levy~Yeyati}},\ and\ \bibinfo {author}
  {\bibfnamefont {E.~J.~H.}\ \bibnamefont {Lee}},\ }\bibfield  {title}
  {\bibinfo {title} {Heat {{Dissipation Mechanisms}} in {{Hybrid
  Superconductor}}--{{Semiconductor Devices Revealed}} by {{Joule
  Spectroscopy}}},\ }\href {https://doi.org/10.1021/acs.nanolett.4c00574}
  {\bibfield  {journal} {\bibinfo  {journal} {Nano Letters}\ }\textbf {\bibinfo
  {volume} {24}},\ \bibinfo {pages} {6488} (\bibinfo {year}
  {2024})}\BibitemShut {NoStop}%
\bibitem [{\citenamefont {Shpagina}\ \emph {et~al.}(2025)\citenamefont
  {Shpagina}, \citenamefont {Tikhonov}, \citenamefont {Ruhstorfer},
  \citenamefont {Koblm{\"u}ller},\ and\ \citenamefont
  {Khrapai}}]{shpagina2025}%
  \BibitemOpen
  \bibfield  {author} {\bibinfo {author} {\bibfnamefont {E.~V.}\ \bibnamefont
  {Shpagina}}, \bibinfo {author} {\bibfnamefont {E.~S.}\ \bibnamefont
  {Tikhonov}}, \bibinfo {author} {\bibfnamefont {D.}~\bibnamefont
  {Ruhstorfer}}, \bibinfo {author} {\bibfnamefont {G.}~\bibnamefont
  {Koblm{\"u}ller}},\ and\ \bibinfo {author} {\bibfnamefont {V.~S.}\
  \bibnamefont {Khrapai}},\ }\bibfield  {title} {\bibinfo {title}
  {Superconducting bistability in floating {{Al}} islands of hybrid
  {{Al}}/{{InAs}} nanowires},\ }\href {https://doi.org/10.1103/l4tq-bxq9}
  {\bibfield  {journal} {\bibinfo  {journal} {Physical Review B}\ }\textbf
  {\bibinfo {volume} {112}},\ \bibinfo {pages} {L140503} (\bibinfo {year}
  {2025})}\BibitemShut {NoStop}%
\bibitem [{\citenamefont {Oliveira}\ \emph {et~al.}(2025)\citenamefont
  {Oliveira}, \citenamefont {Steffensen}, \citenamefont {Iglesias},
  \citenamefont {Gomez}, \citenamefont {Ibabe}, \citenamefont {Kanne},
  \citenamefont {Nygard}, \citenamefont {Aguado}, \citenamefont {Yeyati},\ and\
  \citenamefont {Lee}}]{oliveira2025}%
  \BibitemOpen
  \bibfield  {author} {\bibinfo {author} {\bibfnamefont {G.~M.}\ \bibnamefont
  {Oliveira}}, \bibinfo {author} {\bibfnamefont {G.~O.}\ \bibnamefont
  {Steffensen}}, \bibinfo {author} {\bibfnamefont {I.~C.}\ \bibnamefont
  {Iglesias}}, \bibinfo {author} {\bibfnamefont {M.}~\bibnamefont {Gomez}},
  \bibinfo {author} {\bibfnamefont {A.}~\bibnamefont {Ibabe}}, \bibinfo
  {author} {\bibfnamefont {T.}~\bibnamefont {Kanne}}, \bibinfo {author}
  {\bibfnamefont {J.}~\bibnamefont {Nygard}}, \bibinfo {author} {\bibfnamefont
  {R.}~\bibnamefont {Aguado}}, \bibinfo {author} {\bibfnamefont {A.~L.}\
  \bibnamefont {Yeyati}},\ and\ \bibinfo {author} {\bibfnamefont {E.~J.~H.}\
  \bibnamefont {Lee}},\ }\href {https://arxiv.org/abs/2512.02828} {\bibinfo
  {title} {Anomalous metallic phase and reduced critical current in
  superconducting nanowires due to inverse proximity effect}} (\bibinfo {year}
  {2025}),\ \Eprint {https://arxiv.org/abs/2512.02828} {arXiv:2512.02828
  [cond-mat.mes-hall]} \BibitemShut {NoStop}%
\bibitem [{\citenamefont {Galante-Agero}\ \emph {et~al.}(2026)\citenamefont
  {Galante-Agero}, \citenamefont {Adam}, \citenamefont {Denisov}, \citenamefont
  {Gerber}, \citenamefont {Niese}, \citenamefont {Mestre-Torà}, \citenamefont
  {Perego}, \citenamefont {Richter}, \citenamefont {Taniguchi}, \citenamefont
  {Watanabe}, \citenamefont {Ensslin},\ and\ \citenamefont
  {Ihn}}]{galante-agero2026}%
  \BibitemOpen
  \bibfield  {author} {\bibinfo {author} {\bibfnamefont {C.}~\bibnamefont
  {Galante-Agero}}, \bibinfo {author} {\bibfnamefont {C.}~\bibnamefont {Adam}},
  \bibinfo {author} {\bibfnamefont {A.~O.}\ \bibnamefont {Denisov}}, \bibinfo
  {author} {\bibfnamefont {J.~D.}\ \bibnamefont {Gerber}}, \bibinfo {author}
  {\bibfnamefont {M.}~\bibnamefont {Niese}}, \bibinfo {author} {\bibfnamefont
  {A.}~\bibnamefont {Mestre-Torà}}, \bibinfo {author} {\bibfnamefont
  {M.}~\bibnamefont {Perego}}, \bibinfo {author} {\bibfnamefont
  {J.}~\bibnamefont {Richter}}, \bibinfo {author} {\bibfnamefont
  {T.}~\bibnamefont {Taniguchi}}, \bibinfo {author} {\bibfnamefont
  {K.}~\bibnamefont {Watanabe}}, \bibinfo {author} {\bibfnamefont
  {K.}~\bibnamefont {Ensslin}},\ and\ \bibinfo {author} {\bibfnamefont
  {T.}~\bibnamefont {Ihn}},\ }\href {https://arxiv.org/abs/2608.16711}
  {\bibinfo {title} {Proximity-induced superconductivity in a bilayer graphene
  quantum point contact}} (\bibinfo {year} {2026}),\ \Eprint
  {https://arxiv.org/abs/2608.16711} {arXiv:2608.16711 [cond-mat.mes-hall]}
  \BibitemShut {NoStop}%
\bibitem [{\citenamefont {Courtois}\ \emph {et~al.}(2008)\citenamefont
  {Courtois}, \citenamefont {Meschke}, \citenamefont {Peltonen},\ and\
  \citenamefont {Pekola}}]{courtois2008}%
  \BibitemOpen
  \bibfield  {author} {\bibinfo {author} {\bibfnamefont {H.}~\bibnamefont
  {Courtois}}, \bibinfo {author} {\bibfnamefont {M.}~\bibnamefont {Meschke}},
  \bibinfo {author} {\bibfnamefont {J.~T.}\ \bibnamefont {Peltonen}},\ and\
  \bibinfo {author} {\bibfnamefont {J.~P.}\ \bibnamefont {Pekola}},\ }\bibfield
   {title} {\bibinfo {title} {Origin of {{Hysteresis}} in a {{Proximity
  Josephson Junction}}},\ }\href
  {https://doi.org/10.1103/PhysRevLett.101.067002} {\bibfield  {journal}
  {\bibinfo  {journal} {Physical Review Letters}\ }\textbf {\bibinfo {volume}
  {101}},\ \bibinfo {pages} {067002} (\bibinfo {year} {2008})}\BibitemShut
  {NoStop}%
\bibitem [{\citenamefont {Maisi}\ \emph {et~al.}(2013)\citenamefont {Maisi},
  \citenamefont {Lotkhov}, \citenamefont {Kemppinen}, \citenamefont {Heimes},
  \citenamefont {Muhonen},\ and\ \citenamefont {Pekola}}]{maisi2013}%
  \BibitemOpen
  \bibfield  {author} {\bibinfo {author} {\bibfnamefont {V.~F.}\ \bibnamefont
  {Maisi}}, \bibinfo {author} {\bibfnamefont {S.~V.}\ \bibnamefont {Lotkhov}},
  \bibinfo {author} {\bibfnamefont {A.}~\bibnamefont {Kemppinen}}, \bibinfo
  {author} {\bibfnamefont {A.}~\bibnamefont {Heimes}}, \bibinfo {author}
  {\bibfnamefont {J.~T.}\ \bibnamefont {Muhonen}},\ and\ \bibinfo {author}
  {\bibfnamefont {J.~P.}\ \bibnamefont {Pekola}},\ }\bibfield  {title}
  {\bibinfo {title} {Excitation of {{Single Quasiparticles}} in a {{Small
  Superconducting Al Island Connected}} to {{Normal-Metal Leads}} by {{Tunnel
  Junctions}}},\ }\href {https://doi.org/10.1103/PhysRevLett.111.147001}
  {\bibfield  {journal} {\bibinfo  {journal} {Physical Review Letters}\
  }\textbf {\bibinfo {volume} {111}},\ \bibinfo {pages} {147001} (\bibinfo
  {year} {2013})}\BibitemShut {NoStop}%
\bibitem [{\citenamefont {Andreev}(1964)}]{Andreev}%
  \BibitemOpen
  \bibfield  {author} {\bibinfo {author} {\bibfnamefont {A.~F.}\ \bibnamefont
  {Andreev}},\ }\bibfield  {title} {\bibinfo {title} {Thermal conductivity of
  the intermediate state of superconductors},\ }\href@noop {} {\bibfield
  {journal} {\bibinfo  {journal} {JETP Letters}\ }\textbf {\bibinfo {volume}
  {46}},\ \bibinfo {pages} {1823} (\bibinfo {year} {1964})}\BibitemShut
  {NoStop}%
\bibitem [{\citenamefont {Roukes}\ \emph {et~al.}(1985)\citenamefont {Roukes},
  \citenamefont {Freeman}, \citenamefont {Germain}, \citenamefont
  {Richardson},\ and\ \citenamefont {Ketchen}}]{roukes1985}%
  \BibitemOpen
  \bibfield  {author} {\bibinfo {author} {\bibfnamefont {M.~L.}\ \bibnamefont
  {Roukes}}, \bibinfo {author} {\bibfnamefont {M.~R.}\ \bibnamefont {Freeman}},
  \bibinfo {author} {\bibfnamefont {R.~S.}\ \bibnamefont {Germain}}, \bibinfo
  {author} {\bibfnamefont {R.~C.}\ \bibnamefont {Richardson}},\ and\ \bibinfo
  {author} {\bibfnamefont {M.~B.}\ \bibnamefont {Ketchen}},\ }\bibfield
  {title} {\bibinfo {title} {Hot electrons and energy transport in metals at
  millikelvin temperatures},\ }\href
  {https://doi.org/10.1103/PhysRevLett.55.422} {\bibfield  {journal} {\bibinfo
  {journal} {Physical Review Letters}\ }\textbf {\bibinfo {volume} {55}},\
  \bibinfo {pages} {422} (\bibinfo {year} {1985})}\BibitemShut {NoStop}%
\bibitem [{\citenamefont {Baselmans}\ \emph {et~al.}(2022)\citenamefont
  {Baselmans}, \citenamefont {Facchin}, \citenamefont {Pascual~Laguna},
  \citenamefont {Bueno}, \citenamefont {Thoen}, \citenamefont {Murugesan},
  \citenamefont {Llombart},\ and\ \citenamefont {de~Visser}}]{Baselmans2022}%
  \BibitemOpen
  \bibfield  {author} {\bibinfo {author} {\bibfnamefont {J.~J.~A.}\
  \bibnamefont {Baselmans}}, \bibinfo {author} {\bibfnamefont {F.}~\bibnamefont
  {Facchin}}, \bibinfo {author} {\bibfnamefont {A.}~\bibnamefont
  {Pascual~Laguna}}, \bibinfo {author} {\bibfnamefont {J.}~\bibnamefont
  {Bueno}}, \bibinfo {author} {\bibfnamefont {D.~J.}\ \bibnamefont {Thoen}},
  \bibinfo {author} {\bibfnamefont {V.}~\bibnamefont {Murugesan}}, \bibinfo
  {author} {\bibfnamefont {N.}~\bibnamefont {Llombart}},\ and\ \bibinfo
  {author} {\bibfnamefont {P.~J.}\ \bibnamefont {de~Visser}},\ }\bibfield
  {title} {\bibinfo {title} {Ultra-sensitive {THz} microwave kinetic inductance
  detectors for future space telescopes},\ }\href
  {https://doi.org/10.1051/0004-6361/202243840} {\bibfield  {journal} {\bibinfo
   {journal} {Astron. Astrophys.}\ }\textbf {\bibinfo {volume} {665}},\
  \bibinfo {pages} {A17} (\bibinfo {year} {2022})}\BibitemShut {NoStop}%
\bibitem [{\citenamefont {Pimanov}\ \emph {et~al.}(2025)\citenamefont
  {Pimanov}, \citenamefont {Pankratov}, \citenamefont {Gordeeva}, \citenamefont
  {Chiginev}, \citenamefont {Blagodatkin}, \citenamefont {Revin}, \citenamefont
  {Razov}, \citenamefont {Yu~Safonova}, \citenamefont {Fedotov}, \citenamefont
  {Skorokhodov}, \citenamefont {Orlova}, \citenamefont {Tatarsky},
  \citenamefont {Gusev}, \citenamefont {Trofimov}, \citenamefont {Mumlyakov},\
  and\ \citenamefont {Tarkhov}}]{Pimanov2025}%
  \BibitemOpen
  \bibfield  {author} {\bibinfo {author} {\bibfnamefont {D.~A.}\ \bibnamefont
  {Pimanov}}, \bibinfo {author} {\bibfnamefont {A.~L.}\ \bibnamefont
  {Pankratov}}, \bibinfo {author} {\bibfnamefont {A.~V.}\ \bibnamefont
  {Gordeeva}}, \bibinfo {author} {\bibfnamefont {A.~V.}\ \bibnamefont
  {Chiginev}}, \bibinfo {author} {\bibfnamefont {A.~V.}\ \bibnamefont
  {Blagodatkin}}, \bibinfo {author} {\bibfnamefont {L.~S.}\ \bibnamefont
  {Revin}}, \bibinfo {author} {\bibfnamefont {S.~A.}\ \bibnamefont {Razov}},
  \bibinfo {author} {\bibfnamefont {V.}~\bibnamefont {Yu~Safonova}}, \bibinfo
  {author} {\bibfnamefont {I.~A.}\ \bibnamefont {Fedotov}}, \bibinfo {author}
  {\bibfnamefont {E.~V.}\ \bibnamefont {Skorokhodov}}, \bibinfo {author}
  {\bibfnamefont {A.~N.}\ \bibnamefont {Orlova}}, \bibinfo {author}
  {\bibfnamefont {D.~A.}\ \bibnamefont {Tatarsky}}, \bibinfo {author}
  {\bibfnamefont {N.~S.}\ \bibnamefont {Gusev}}, \bibinfo {author}
  {\bibfnamefont {I.~V.}\ \bibnamefont {Trofimov}}, \bibinfo {author}
  {\bibfnamefont {A.~M.}\ \bibnamefont {Mumlyakov}},\ and\ \bibinfo {author}
  {\bibfnamefont {M.~A.}\ \bibnamefont {Tarkhov}},\ }\bibfield  {title}
  {\bibinfo {title} {Response of a cold-electron bolometer in a coplanar
  antenna system},\ }\href {https://doi.org/10.1088/1361-6668/adb942}
  {\bibfield  {journal} {\bibinfo  {journal} {Supercond. Sci. Technol.}\
  }\textbf {\bibinfo {volume} {38}},\ \bibinfo {pages} {035026} (\bibinfo
  {year} {2025})}\BibitemShut {NoStop}%
\bibitem [{\citenamefont {Hoffmann}\ \emph {et~al.}(2009)\citenamefont
  {Hoffmann}, \citenamefont {Nilsson}, \citenamefont {Matthews}, \citenamefont
  {Nakpathomkun}, \citenamefont {Persson}, \citenamefont {Samuelson},\ and\
  \citenamefont {Linke}}]{hoffmann2009}%
  \BibitemOpen
  \bibfield  {author} {\bibinfo {author} {\bibfnamefont {E.~A.}\ \bibnamefont
  {Hoffmann}}, \bibinfo {author} {\bibfnamefont {H.~A.}\ \bibnamefont
  {Nilsson}}, \bibinfo {author} {\bibfnamefont {J.~E.}\ \bibnamefont
  {Matthews}}, \bibinfo {author} {\bibfnamefont {N.}~\bibnamefont
  {Nakpathomkun}}, \bibinfo {author} {\bibfnamefont {A.~I.}\ \bibnamefont
  {Persson}}, \bibinfo {author} {\bibfnamefont {L.}~\bibnamefont {Samuelson}},\
  and\ \bibinfo {author} {\bibfnamefont {H.}~\bibnamefont {Linke}},\ }\bibfield
   {title} {\bibinfo {title} {Measuring {{Temperature Gradients}} over
  {{Nanometer Length Scales}}},\ }\href {https://doi.org/10.1021/nl8034042}
  {\bibfield  {journal} {\bibinfo  {journal} {Nano Letters}\ }\textbf {\bibinfo
  {volume} {9}},\ \bibinfo {pages} {779} (\bibinfo {year} {2009})}\BibitemShut
  {NoStop}%
\bibitem [{\citenamefont {Artemenko}\ \emph {et~al.}(1979)\citenamefont
  {Artemenko}, \citenamefont {Volkov},\ and\ \citenamefont
  {Zaitsev}}]{Artemenko1979}%
  \BibitemOpen
  \bibfield  {author} {\bibinfo {author} {\bibfnamefont {S.}~\bibnamefont
  {Artemenko}}, \bibinfo {author} {\bibfnamefont {A.}~\bibnamefont {Volkov}},\
  and\ \bibinfo {author} {\bibfnamefont {A.}~\bibnamefont {Zaitsev}},\
  }\bibfield  {title} {\bibinfo {title} {On the excess current in microbridges
  {{S}}-c-{{S}} and {{S}}-c-{{N}}},\ }\href
  {https://doi.org/10.1016/0038-1098(79)90044-9} {\bibfield  {journal}
  {\bibinfo  {journal} {Solid State Communications}\ }\textbf {\bibinfo
  {volume} {30}},\ \bibinfo {pages} {771} (\bibinfo {year} {1979})}\BibitemShut
  {NoStop}%
\bibitem [{\citenamefont {Volkov}\ \emph {et~al.}(1996)\citenamefont {Volkov},
  \citenamefont {Allsopp},\ and\ \citenamefont {Lambert}}]{Volkov1996}%
  \BibitemOpen
  \bibfield  {author} {\bibinfo {author} {\bibfnamefont {A.}~\bibnamefont
  {Volkov}}, \bibinfo {author} {\bibfnamefont {N.}~\bibnamefont {Allsopp}},\
  and\ \bibinfo {author} {\bibfnamefont {C.~J.}\ \bibnamefont {Lambert}},\
  }\bibfield  {title} {\bibinfo {title} {Crossover from mesoscopic to classical
  proximity effects, induced by particle - hole symmetry breaking in andreev
  interferometers},\ }\href {https://doi.org/10.1088/0953-8984/8/4/001}
  {\bibfield  {journal} {\bibinfo  {journal} {Journal of Physics: Condensed
  Matter}\ }\textbf {\bibinfo {volume} {8}},\ \bibinfo {pages} {L45} (\bibinfo
  {year} {1996})}\BibitemShut {NoStop}%
\bibitem [{\citenamefont {Nazarov}\ and\ \citenamefont
  {Stoof}(1996)}]{Nazarov1996}%
  \BibitemOpen
  \bibfield  {author} {\bibinfo {author} {\bibfnamefont {Y.~V.}\ \bibnamefont
  {Nazarov}}\ and\ \bibinfo {author} {\bibfnamefont {T.~H.}\ \bibnamefont
  {Stoof}},\ }\bibfield  {title} {\bibinfo {title} {Diffusive conductors as
  andreev interferometers},\ }\href
  {https://doi.org/10.1103/physrevlett.76.823} {\bibfield  {journal} {\bibinfo
  {journal} {Physical Review Letters}\ }\textbf {\bibinfo {volume} {76}},\
  \bibinfo {pages} {823} (\bibinfo {year} {1996})}\BibitemShut {NoStop}%
\bibitem [{\citenamefont {Lachenmann}\ \emph {et~al.}(1997)\citenamefont
  {Lachenmann}, \citenamefont {Friedrich}, \citenamefont {F\"{o}rster},
  \citenamefont {Uhlisch},\ and\ \citenamefont {Golubov}}]{Lachenmann1997}%
  \BibitemOpen
  \bibfield  {author} {\bibinfo {author} {\bibfnamefont {S.~G.}\ \bibnamefont
  {Lachenmann}}, \bibinfo {author} {\bibfnamefont {I.}~\bibnamefont
  {Friedrich}}, \bibinfo {author} {\bibfnamefont {A.}~\bibnamefont
  {F\"{o}rster}}, \bibinfo {author} {\bibfnamefont {D.}~\bibnamefont
  {Uhlisch}},\ and\ \bibinfo {author} {\bibfnamefont {A.~A.}\ \bibnamefont
  {Golubov}},\ }\bibfield  {title} {\bibinfo {title} {Charge transport in
  superconductor/semiconductor/normal-conductor step junctions},\ }\href
  {https://doi.org/10.1103/physrevb.56.14108} {\bibfield  {journal} {\bibinfo
  {journal} {Physical Review B}\ }\textbf {\bibinfo {volume} {56}},\ \bibinfo
  {pages} {14108} (\bibinfo {year} {1997})}\BibitemShut {NoStop}%
\bibitem [{\citenamefont {Golubov}\ \emph {et~al.}(1997)\citenamefont
  {Golubov}, \citenamefont {Wilhelm},\ and\ \citenamefont
  {Zaikin}}]{Golubov1997}%
  \BibitemOpen
  \bibfield  {author} {\bibinfo {author} {\bibfnamefont {A.~A.}\ \bibnamefont
  {Golubov}}, \bibinfo {author} {\bibfnamefont {F.~K.}\ \bibnamefont
  {Wilhelm}},\ and\ \bibinfo {author} {\bibfnamefont {A.~D.}\ \bibnamefont
  {Zaikin}},\ }\bibfield  {title} {\bibinfo {title} {Coherent charge transport
  in metallic proximity structures},\ }\href
  {https://doi.org/10.1103/physrevb.55.1123} {\bibfield  {journal} {\bibinfo
  {journal} {Physical Review B}\ }\textbf {\bibinfo {volume} {55}},\ \bibinfo
  {pages} {1123} (\bibinfo {year} {1997})}\BibitemShut {NoStop}%
\bibitem [{\citenamefont {Kozhevnikov}\ \emph {et~al.}(2000)\citenamefont
  {Kozhevnikov}, \citenamefont {Schoelkopf},\ and\ \citenamefont
  {Prober}}]{Kozhevnikov2000}%
  \BibitemOpen
  \bibfield  {author} {\bibinfo {author} {\bibfnamefont {A.~A.}\ \bibnamefont
  {Kozhevnikov}}, \bibinfo {author} {\bibfnamefont {R.~J.}\ \bibnamefont
  {Schoelkopf}},\ and\ \bibinfo {author} {\bibfnamefont {D.~E.}\ \bibnamefont
  {Prober}},\ }\bibfield  {title} {\bibinfo {title} {Observation of
  photon-assisted noise in a diffusive normal metal superconductor junction},\
  }\href {https://doi.org/10.1103/PhysRevLett.84.3398} {\bibfield  {journal}
  {\bibinfo  {journal} {Phys. Rev. Lett.}\ }\textbf {\bibinfo {volume} {84}},\
  \bibinfo {pages} {3398} (\bibinfo {year} {2000})}\BibitemShut {NoStop}%
\bibitem [{\citenamefont {Bubis}\ \emph {et~al.}(2017)\citenamefont {Bubis},
  \citenamefont {Denisov}, \citenamefont {Piatrusha}, \citenamefont {Batov},
  \citenamefont {Khrapai}, \citenamefont {Becker}, \citenamefont {Treu},
  \citenamefont {Ruhstorfer},\ and\ \citenamefont
  {Koblm{\"u}ller}}]{bubis2017}%
  \BibitemOpen
  \bibfield  {author} {\bibinfo {author} {\bibfnamefont {A.~V.}\ \bibnamefont
  {Bubis}}, \bibinfo {author} {\bibfnamefont {A.~O.}\ \bibnamefont {Denisov}},
  \bibinfo {author} {\bibfnamefont {S.~U.}\ \bibnamefont {Piatrusha}}, \bibinfo
  {author} {\bibfnamefont {I.~E.}\ \bibnamefont {Batov}}, \bibinfo {author}
  {\bibfnamefont {V.~S.}\ \bibnamefont {Khrapai}}, \bibinfo {author}
  {\bibfnamefont {J.}~\bibnamefont {Becker}}, \bibinfo {author} {\bibfnamefont
  {J.}~\bibnamefont {Treu}}, \bibinfo {author} {\bibfnamefont {D.}~\bibnamefont
  {Ruhstorfer}},\ and\ \bibinfo {author} {\bibfnamefont {G.}~\bibnamefont
  {Koblm{\"u}ller}},\ }\bibfield  {title} {\bibinfo {title} {Proximity effect
  and interface transparency in {{Al}}/ {{InAs-nanowire}}/ {{Al}} diffusive
  junctions},\ }\href {https://doi.org/10.1088/1361-6641/aa7eef} {\bibfield
  {journal} {\bibinfo  {journal} {Semicond. Sci. Technol.}\ }\textbf {\bibinfo
  {volume} {32}},\ \bibinfo {pages} {094007} (\bibinfo {year}
  {2017})}\BibitemShut {NoStop}%
\bibitem [{\citenamefont {{E S Tikhonov}}\ \emph {et~al.}(2016)\citenamefont
  {{E S Tikhonov}}, \citenamefont {{D V Shovkun}}, \citenamefont {{D
  Ercolani}}, \citenamefont {{F Rossella}}, \citenamefont {{M Rocci}},
  \citenamefont {{L Sorba}}, \citenamefont {{S Roddaro}},\ and\ \citenamefont
  {{V S Khrapai}}}]{tikhonov2016a}%
  \BibitemOpen
  \bibfield  {author} {\bibinfo {author} {\bibnamefont {{E S Tikhonov}}},
  \bibinfo {author} {\bibnamefont {{D V Shovkun}}}, \bibinfo {author}
  {\bibnamefont {{D Ercolani}}}, \bibinfo {author} {\bibnamefont {{F
  Rossella}}}, \bibinfo {author} {\bibnamefont {{M Rocci}}}, \bibinfo {author}
  {\bibnamefont {{L Sorba}}}, \bibinfo {author} {\bibnamefont {{S Roddaro}}},\
  and\ \bibinfo {author} {\bibnamefont {{V S Khrapai}}},\ }\bibfield  {title}
  {\bibinfo {title} {Local noise in a diffusive conductor},\ }\href
  {https://doi.org/10.1038/srep30621} {\bibfield  {journal} {\bibinfo
  {journal} {Scientific Reports}\ }\textbf {\bibinfo {volume} {6}},\ \bibinfo
  {pages} {30621} (\bibinfo {year} {2016})}\BibitemShut {NoStop}%
\bibitem [{\citenamefont {Tikhonov}\ \emph {et~al.}(2016)\citenamefont
  {Tikhonov}, \citenamefont {Shovkun}, \citenamefont {Ercolani}, \citenamefont
  {Rossella}, \citenamefont {Rocci}, \citenamefont {Sorba}, \citenamefont
  {Roddaro},\ and\ \citenamefont {Khrapai}}]{tikhonov2016b}%
  \BibitemOpen
  \bibfield  {author} {\bibinfo {author} {\bibfnamefont {E.~S.}\ \bibnamefont
  {Tikhonov}}, \bibinfo {author} {\bibfnamefont {D.~V.}\ \bibnamefont
  {Shovkun}}, \bibinfo {author} {\bibfnamefont {D.}~\bibnamefont {Ercolani}},
  \bibinfo {author} {\bibfnamefont {F.}~\bibnamefont {Rossella}}, \bibinfo
  {author} {\bibfnamefont {M.}~\bibnamefont {Rocci}}, \bibinfo {author}
  {\bibfnamefont {L.}~\bibnamefont {Sorba}}, \bibinfo {author} {\bibfnamefont
  {S.}~\bibnamefont {Roddaro}},\ and\ \bibinfo {author} {\bibfnamefont {V.~S.}\
  \bibnamefont {Khrapai}},\ }\bibfield  {title} {\bibinfo {title} {Noise
  thermometry applied to thermoelectric measurements in {{InAs}} nanowires},\
  }\href {https://doi.org/10.1088/0268-1242/31/10/104001} {\bibfield  {journal}
  {\bibinfo  {journal} {Semicond. Sci. Technol.}\ }\textbf {\bibinfo {volume}
  {31}},\ \bibinfo {pages} {104001} (\bibinfo {year} {2016})}\BibitemShut
  {NoStop}%
\bibitem [{\citenamefont {Baeva}\ \emph {et~al.}(2021)\citenamefont {Baeva},
  \citenamefont {Titova}, \citenamefont {Veyrat}, \citenamefont
  {Sac{\'e}p{\'e}}, \citenamefont {Semenov}, \citenamefont {Goltsman},
  \citenamefont {Kardakova},\ and\ \citenamefont {Khrapai}}]{baeva2021}%
  \BibitemOpen
  \bibfield  {author} {\bibinfo {author} {\bibfnamefont {E.~M.}\ \bibnamefont
  {Baeva}}, \bibinfo {author} {\bibfnamefont {N.~A.}\ \bibnamefont {Titova}},
  \bibinfo {author} {\bibfnamefont {L.}~\bibnamefont {Veyrat}}, \bibinfo
  {author} {\bibfnamefont {B.}~\bibnamefont {Sac{\'e}p{\'e}}}, \bibinfo
  {author} {\bibfnamefont {A.~V.}\ \bibnamefont {Semenov}}, \bibinfo {author}
  {\bibfnamefont {G.~N.}\ \bibnamefont {Goltsman}}, \bibinfo {author}
  {\bibfnamefont {A.~I.}\ \bibnamefont {Kardakova}},\ and\ \bibinfo {author}
  {\bibfnamefont {{\relax Vadim}.~S.}\ \bibnamefont {Khrapai}},\ }\bibfield
  {title} {\bibinfo {title} {Thermal {{Relaxation}} in {{Metal Films Limited}}
  by {{Diffuson Lattice Excitations}} of {{Amorphous Substrates}}},\ }\href
  {https://doi.org/10.1103/PhysRevApplied.15.054014} {\bibfield  {journal}
  {\bibinfo  {journal} {Physical Review Applied}\ }\textbf {\bibinfo {volume}
  {15}},\ \bibinfo {pages} {054014} (\bibinfo {year} {2021})}\BibitemShut
  {NoStop}%
\bibitem [{\citenamefont {Swartz}\ and\ \citenamefont
  {Pohl}(1989)}]{swartz1989}%
  \BibitemOpen
  \bibfield  {author} {\bibinfo {author} {\bibfnamefont {E.~T.}\ \bibnamefont
  {Swartz}}\ and\ \bibinfo {author} {\bibfnamefont {R.~O.}\ \bibnamefont
  {Pohl}},\ }\bibfield  {title} {\bibinfo {title} {Thermal boundary
  resistance},\ }\href {https://doi.org/10.1103/RevModPhys.61.605} {\bibfield
  {journal} {\bibinfo  {journal} {Reviews of Modern Physics}\ }\textbf
  {\bibinfo {volume} {61}},\ \bibinfo {pages} {605} (\bibinfo {year}
  {1989})}\BibitemShut {NoStop}%
\bibitem [{Note1()}]{Note1}%
  \BibitemOpen
  \bibinfo {note} {A relatively weak thermal coupling of the S-island to the NW
  contact pads as compared to the $^3$He bath makes any possible asymmetry of
  the pad temperatures for $I_\protect \mathrm {H}\protect \neq 0$
  unimportant.}\BibitemShut {Stop}%
\bibitem [{\citenamefont {Peshkov}(1968)}]{peshkov1968}%
  \BibitemOpen
  \bibfield  {author} {\bibinfo {author} {\bibfnamefont {V.~N.}\ \bibnamefont
  {Peshkov}},\ }\bibfield  {title} {\bibinfo {title} {{{Properties of
  He}}{\textsuperscript{3}} {{and of its solutions in
  He}}{\textsuperscript{4}}},\ }\href
  {https://doi.org/10.1070/PU1968v011n02ABEH003812} {\bibfield  {journal}
  {\bibinfo  {journal} {Soviet Physics Uspekhi}\ }\textbf {\bibinfo {volume}
  {11}},\ \bibinfo {pages} {209} (\bibinfo {year} {1968})}\BibitemShut
  {NoStop}%
\bibitem [{\citenamefont {Behnia}\ and\ \citenamefont
  {Trachenko}(2024)}]{behnia2024}%
  \BibitemOpen
  \bibfield  {author} {\bibinfo {author} {\bibfnamefont {K.}~\bibnamefont
  {Behnia}}\ and\ \bibinfo {author} {\bibfnamefont {K.}~\bibnamefont
  {Trachenko}},\ }\bibfield  {title} {\bibinfo {title} {How heat propagates in
  liquid {{3He}}},\ }\href {https://doi.org/10.1038/s41467-024-46079-0}
  {\bibfield  {journal} {\bibinfo  {journal} {Nature Communications}\ }\textbf
  {\bibinfo {volume} {15}},\ \bibinfo {pages} {1771} (\bibinfo {year}
  {2024})}\BibitemShut {NoStop}%
\bibitem [{\citenamefont {Anderson}\ and\ \citenamefont
  {Johnson}(1972)}]{anderson1972}%
  \BibitemOpen
  \bibfield  {author} {\bibinfo {author} {\bibfnamefont {A.~C.}\ \bibnamefont
  {Anderson}}\ and\ \bibinfo {author} {\bibfnamefont {W.~L.}\ \bibnamefont
  {Johnson}},\ }\bibfield  {title} {\bibinfo {title} {The {{Kapitza}}
  resistance between copper and {\textsuperscript{3}}{{He}}},\ }\href
  {https://doi.org/10.1007/BF00629118} {\bibfield  {journal} {\bibinfo
  {journal} {Journal of Low Temperature Physics}\ }\textbf {\bibinfo {volume}
  {7}},\ \bibinfo {pages} {1} (\bibinfo {year} {1972})}\BibitemShut {NoStop}%
\bibitem [{\citenamefont {Keizer}\ \emph {et~al.}(2006)\citenamefont {Keizer},
  \citenamefont {Flokstra}, \citenamefont {Aarts},\ and\ \citenamefont
  {Klapwijk}}]{keizer2006}%
  \BibitemOpen
  \bibfield  {author} {\bibinfo {author} {\bibfnamefont {R.~S.}\ \bibnamefont
  {Keizer}}, \bibinfo {author} {\bibfnamefont {M.~G.}\ \bibnamefont
  {Flokstra}}, \bibinfo {author} {\bibfnamefont {J.}~\bibnamefont {Aarts}},\
  and\ \bibinfo {author} {\bibfnamefont {T.~M.}\ \bibnamefont {Klapwijk}},\
  }\bibfield  {title} {\bibinfo {title} {Critical {{Voltage}} of a {{Mesoscopic
  Superconductor}}},\ }\href {https://doi.org/10.1103/PhysRevLett.96.147002}
  {\bibfield  {journal} {\bibinfo  {journal} {Physical Review Letters}\
  }\textbf {\bibinfo {volume} {96}},\ \bibinfo {pages} {147002} (\bibinfo
  {year} {2006})}\BibitemShut {NoStop}%
\bibitem [{\citenamefont {Snyman}\ and\ \citenamefont
  {Nazarov}(2009)}]{snyman2009}%
  \BibitemOpen
  \bibfield  {author} {\bibinfo {author} {\bibfnamefont {I.}~\bibnamefont
  {Snyman}}\ and\ \bibinfo {author} {\bibfnamefont {{\relax Yu}.~V.}\
  \bibnamefont {Nazarov}},\ }\bibfield  {title} {\bibinfo {title} {Bistability
  in voltage-biased normal-metal/ insulator/ superconductor/ insulator/
  normal-metal structures},\ }\href
  {https://doi.org/10.1103/PhysRevB.79.014510} {\bibfield  {journal} {\bibinfo
  {journal} {Physical Review B}\ }\textbf {\bibinfo {volume} {79}},\ \bibinfo
  {pages} {014510} (\bibinfo {year} {2009})}\BibitemShut {NoStop}%
\bibitem [{\citenamefont {Kawamura}\ \emph {et~al.}(2024)\citenamefont
  {Kawamura}, \citenamefont {Ohashi},\ and\ \citenamefont
  {Stoof}}]{kawamura2024}%
  \BibitemOpen
  \bibfield  {author} {\bibinfo {author} {\bibfnamefont {T.}~\bibnamefont
  {Kawamura}}, \bibinfo {author} {\bibfnamefont {Y.}~\bibnamefont {Ohashi}},\
  and\ \bibinfo {author} {\bibfnamefont {H.~T.~C.}\ \bibnamefont {Stoof}},\
  }\bibfield  {title} {\bibinfo {title} {Emergence of
  {{Larkin-Ovchinnikov-type}} superconducting state in a voltage-driven
  superconductor},\ }\href {https://doi.org/10.1103/PhysRevB.109.104502}
  {\bibfield  {journal} {\bibinfo  {journal} {Physical Review B}\ }\textbf
  {\bibinfo {volume} {109}},\ \bibinfo {pages} {104502} (\bibinfo {year}
  {2024})}\BibitemShut {NoStop}%
\bibitem [{\citenamefont {Kawamura}\ and\ \citenamefont
  {Ohashi}(2025)}]{kawamura2025a}%
  \BibitemOpen
  \bibfield  {author} {\bibinfo {author} {\bibfnamefont {T.}~\bibnamefont
  {Kawamura}}\ and\ \bibinfo {author} {\bibfnamefont {Y.}~\bibnamefont
  {Ohashi}},\ }\bibfield  {title} {\bibinfo {title} {Engineering
  {{Nonequilibrium Superconducting Phases}} in a {{Voltage}}-{{Driven
  Superconductor Under}} an {{External Magnetic Field}}},\ }\href
  {https://doi.org/10.1002/andp.202500102} {\bibfield  {journal} {\bibinfo
  {journal} {Annalen der Physik}\ }\textbf {\bibinfo {volume} {537}},\ \bibinfo
  {pages} {e2500102} (\bibinfo {year} {2025})}\BibitemShut {NoStop}%
\bibitem [{\citenamefont {Kawamura}\ and\ \citenamefont
  {Ohashi}(2026)}]{kawamura2026}%
  \BibitemOpen
  \bibfield  {author} {\bibinfo {author} {\bibfnamefont {T.}~\bibnamefont
  {Kawamura}}\ and\ \bibinfo {author} {\bibfnamefont {Y.}~\bibnamefont
  {Ohashi}},\ }\bibfield  {title} {\bibinfo {title} {Emergence of
  charge-imbalanced bcs state and suppression of the nonequilibrium
  {{Fulde-Ferrell-Larkin-Ovchinnikov}} state in asymmetric nsn junctions},\
  }\href {https://doi.org/10.1103/1q8c-yhd1} {\bibfield  {journal} {\bibinfo
  {journal} {Phys. Rev. B}\ }\textbf {\bibinfo {volume} {113}},\ \bibinfo
  {pages} {214504} (\bibinfo {year} {2026})}\BibitemShut {NoStop}%
\end{thebibliography}

%

\end{document}


\title{Supplementary Material to the article \glqq Thermal bottleneck in a freely suspended superconducting island on InAs nanowire\grqq}


\maketitle
\section{Devices}
\begin{figure}[t!]
	\begin{center}
		\vspace{0mm}
		\includegraphics[width=1\linewidth]{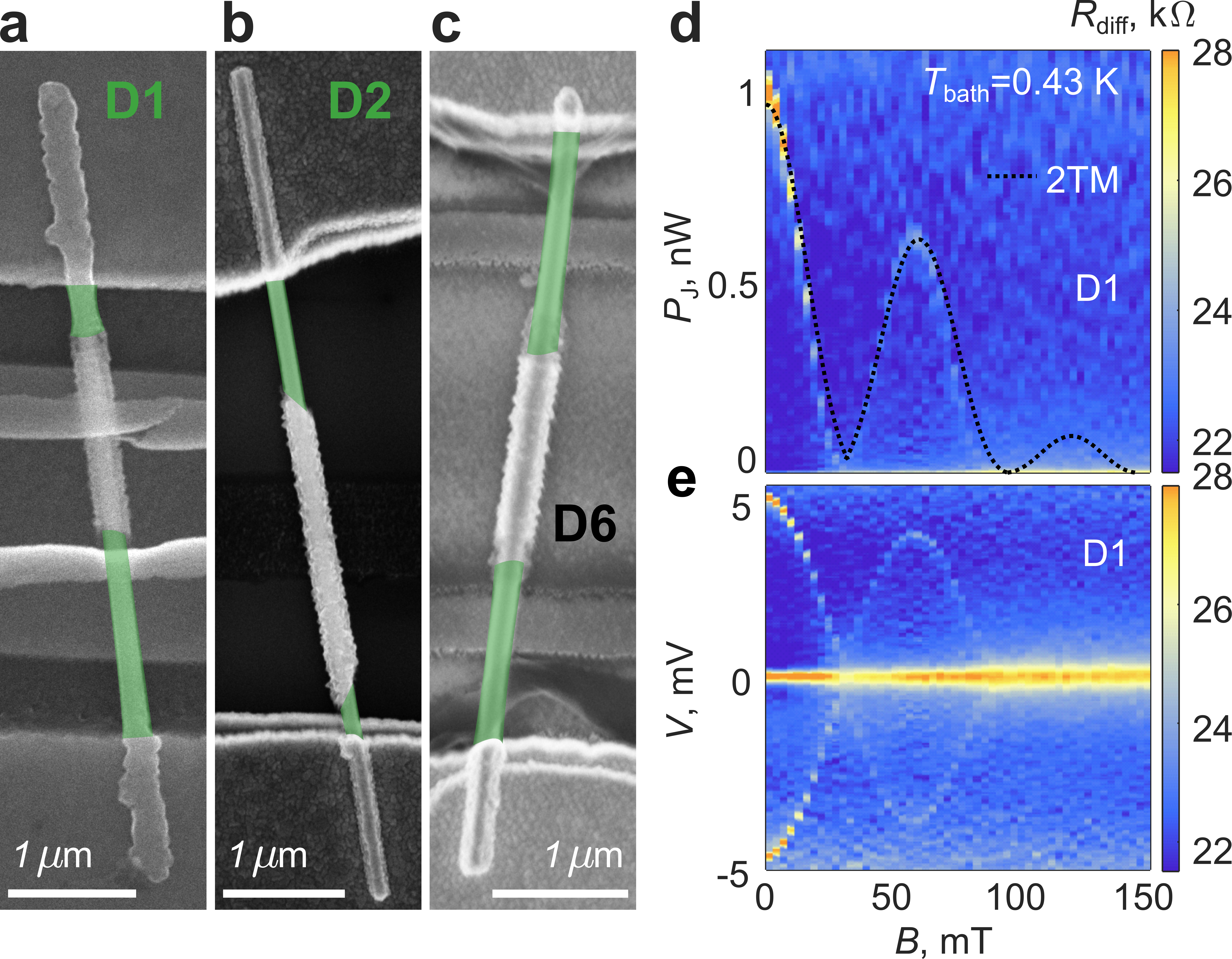}
	\end{center}
	\caption{ a-c: Scanning electron micrographs of the devices D1, D2 and D6. Regions with an exposed InAs core are false-colored in green.
	d-e: Color-scale plots of $R_\text{diff} = dV/dI_\text{NW}$ vs. parallel field $B$ and Joule power $P_\text{J}$ (panel c), bias voltage $V$ (panel d) for the device D1. The dotted line is the 2TM fit. The heaters were off throughout the measurement.  } 
	\label{FigS1}
\end{figure}

Details of the molecular beam epitaxy growth of the nanowires (NWs) are given in~\cite{shpagina2024}. 
The devices studied in the main text are distributed across three substrates: 
D1 on one substrate; D2, D3, D4 on another; and D5, D6 on the third. 
Several NWs have appeared in our previous works: D5 was labeled IIE in~\cite{shpagina2024} and D2 in~\cite{shpagina2025}; 
D3 and D4 correspond to D1 and D5 in~\cite{shpagina2025}, respectively. 
For the previously unreported devices, we provide additional images in Fig.~\ref{FigS1}a-c. For device D1, the top contact pads were fabricated by optical lithography and magnetron sputtering of aluminum (thickness 250~nm). Bright stripes crossing the NW, visible in Fig.~\ref{FigS1}a and Fig.~1b of the main text, are non-conducting residues of baked resist. These residues were removed by ultrasonic cleaning in acetone before the measurements.
For all other devices (D2--D6), the top contact pads were fabricated by thermal evaporation of Cr/Au (2.5 nm/250~nm).

From the fits of the critical Joule power as a function of the magnetic field we obtain the InAs core radius \(\rho_{\text{i}}\) in each device. The Al shell thickness is taken as \(t = 42\)~nm and the superconducting coherence length as \(\xi_0 = 156\)~nm in all devices. The length $L$ of the Al island (S-island) was estimated using scanning electron microscope images.
Table~\ref{table} summarizes the device parameters and the extracted environmental cooling parameter $\Sigma_{\text{env}}$ for the three-temperature model (3TM) presented in Figs.~2 and Figs.~3 of the main text.
The electron-phonon cooling power is fixed at $\Sigma_{\mathrm{e-ph}} = 5.2~\mathrm{nW}\,\mu\mathrm{m}^{-3}\,\mathrm{K}^{-5}$ in all devices.

\begin{table}[h!]
	\centering 
	\caption{ Summary of devices and 3TM parameters. }          
	\begin{tabular}{|c|c|c|c|}\hline 
		Device &  $\rho_\text{i}$, nm & $L$, $\mu$m  & $\Sigma_\text{env}$, $\text{nW}/\text{K}$ \\\hline
		D1 & 82 & 1.57	 &  1.21-2.01 \\
		D2 & 82 & 2.2	 &  0.34 \\
		D3 & 65 & 2.2	 &  0.9 \\
		D4 & 68 & 2.2	 &  0.62 \\
		D5 & 76 & 1.55    &  0.37 \\
		D6 & 82 & 1.5   &  0.61 \\ \hline
	\end{tabular} 
	\label{table}
\end{table}

For device D1 the differential resistance measurements in parallel field (Fig.~\ref{FigS1}e) show a third Little-Parks lobe indicating good alignment of the NW with the $B$-field. We chose the misorientation angle $\alpha = 0$ in this device. The two-temperature model (2TM) fit shown by the dotted line in Fig.~\ref{FigS1}d gives an underestimated value of the electron-phonon cooling power $\Sigma^\text{2TM}_{\mathrm{e-ph}} = 4~\mathrm{nW}\,\mu\mathrm{m}^{-3}\,\mathrm{K}^{-5}$.

\section{Compensation scheme for contact heaters}

\begin{figure}[t!]
	\begin{center}
		\vspace{0mm}
		\includegraphics[width=1\linewidth]{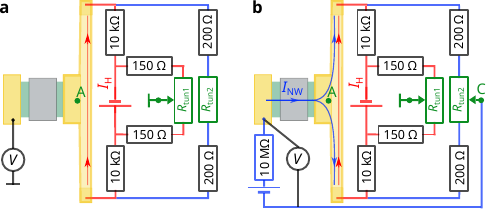}
	\end{center}
	\caption{Sketch of the device layout and the complete electrical circuit for the contact heater, along with the tuning procedure for the compensation circuit. The resistor values correspond to those used in experiments with the device D1.}
	\label{FigS2}
\end{figure}

In the contact heater configuration care was taken to compensate for possible DC (direct current) offsets in the NW circuit in the presence of a finite heater current ($I_\mathrm{H}$). Typical values of the $I_\mathrm{H}$  are about three orders of magnitude higher than the NW transport current ($I_\mathrm{NW}$). In order to disentangle the two currents we employed a compensation circuit shown in Fig.~\ref{FigS2} (this particular  circuit was used in experiments with D1). Apart from the indicated resistances, the circuit contains two 1 k$\Omega$ potentiometers Rtun1 and Rtun2 adjusted to provide virtual grounding of the NW circuit. The tuning procedure is performed in two steps. In the first step, see Fig.~\ref{FigS2}a, we disconnect the NW wiring  and adjust the heater circuit to make sure that the point A is maintained at a virtual ground potential at any $I_\text{H}$. For this, we adjust \(R_{\text{tun1}}\) while the heater current is applied untill  the voltage measured between the A and the ground is nulled. In the second step, see Fig.~\ref{FigS2}b, we equalize the potential of the point C to that of A and the virtual ground when the NW wiring is connected. This is achieved by adjusting Rtun2 so that the voltage difference A-C nulls for $I_\text{NW} =0$ and for any finite $I_\text{H}$.
This procedure is performed before each experiment with the contact heater. As a result of such compensation the measured $P_\text{J}$ is independent of the polarity $I_\text{H}$, see the Figs.1d and 2c in the main text.

\section{Johnson noise thermometry }

\begin{figure}[t!]
	\begin{center}
		\vspace{0mm}
		\includegraphics[width=1\linewidth]{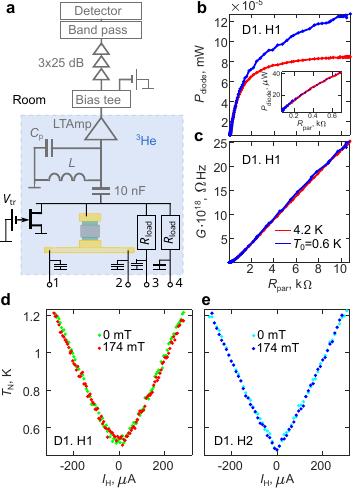}
	\end{center}
	\caption{ a: Noise measurement scheme.
		b: Detector power at two temperatures (4.2~K and 0.6~K) for noise calibration for the D1 device (offset subtracted). The higher temperature dataset (red circles) is divided by a factor of $7 = 4.2/0.6$.  
		c: Current noise gain of the circuit at the same two temperatures as a function of the total load resistance $R_\text{par} = (1/R_\text{NW} + 1/R_\text{tr} + 2/R_\text{load} )^{-1} $.
		d-e: Noise temperature for device D1 in $B=0$ and $B=174$~mT. } 
	\label{FigS3}
\end{figure}

We employ noise thermometry to convert the heater current $I_{\mathrm{H}}$ to the effective temperature rise of the NW, similar to previous experiments  \cite{Tikhonov2016a,Tikhonov2016b}.  The measurement setup is shown schematically in Fig.~\ref{FigS3}a. The NW device is connected to a low-temperature resonant tank circuit (inductance $L \approx 3$~$\mu$H, total capacitance $C_p \approx 20$~pF dominated by a coaxial cable) with a resonance frequency of about 20~MHz. The tank circuit is placed at the input of a home-made low-noise cryogenic amplifier (LTAmp) with a typical input current noise of $S_\text{I}^\text{Amp} \sim 2\cdot 10^{-27}$~A$^2$/Hz. The signal is further amplified by a cascade of room temperature low noise amplifiers (75~dB gain) followed by a bandpass filter and a power detector. 

Two load resistors $R_\text{load}=39\,\mathrm{k \Omega}$ are connected in parallel to the tank circuit and used for quasi-four-terminal measurements scheme (outputs 1-4 in Fig.~\ref{FigS3}a). The gain and the input current noise are calibrated with the help of commercial pHEMT transistor (ATF-55143) in parallel to the device. During all other measurements the transistor gate voltage ($V_\text{tr}$) is zero, which corresponds to pinch-off. 

\subsection*{Detector signal}

Total input impedance of the circuit is 
$Z_{\text{total}}^{-1}(\omega)=1/R_{\text{par}} + 1/(i\omega L) + i\omega C_p$ where $i$ is the imaginary unit and $R_\text{par}^{-1} =  R_\text{NW}^{-1} + R_\text{tr}^{-1} + 2 R_\text{load}^{-1}$ with the NW resistance ($R_\text{NW}$), transistor resistance ($R_\text{tr}$). 
The resistive part generates thermal Johnson–Nyquist noise.
This thermal noise through the input impedance producing a voltage noise spectral density:
\begin{equation}
	S_V^{\text{th}}(\omega)= 4k_\text{B}T \cdot \text{Re}  (Z_{\text{total}}(\omega)) =\frac{4k_\text{B} T_0}{R_{\text{par}}}\,|Z_{\text{total}}(\omega)|^2
\end{equation}
The LTAmp adds current noise $S_I^{\text{Amp}}$, giving an additional voltage contribution $S_I^{\text{Amp}}|Z_{\text{total}}(\omega)|^2$. 
The total voltage noise is then
\begin{equation}
	S_V^{\text{total}}(\omega)=\left(\frac{4k_B T_0}{R_{\text{par}}}+S_I^{\text{Amp}}\right)|Z_{\text{total}}(\omega)|^2.
\end{equation}

The signal passes through a voltage gain $G_0$ and a bandpass filter $F(\omega)$. The power detector integrates over frequency:
\begin{equation}
	P=P_0 + \left(\frac{4k_B T_0}{R_{\text{par}}}+S_I^{\text{Amp}}\right) G_0^2 \int_0^\infty |F(\omega)|^2|Z_{\text{total}}(\omega)|^2\,\frac{d\omega}{2\pi}
\end{equation}
$P_0$ is the constant offset of the detector output signal which is present even in the absence of any input noise. We define the current noise gain of the entire circuit as $G =  G_0^2 \int_0^\infty |F(\omega)|^2|Z_{\text{total}}(\omega)|^2\,\frac{d\omega}{2\pi}$.

\subsection*{Calibration and shot noise measurement}
The calibration relies on two known bath temperatures: $4.2$~K and the post-condensation temperature $T_0$ of the $^3$He, which is the actual measurement temperature. The detector power (with the offset subtracted) as a function of $R_\text{par}$ at two temperatures is shown in Fig.~\ref{FigS3}b for device D1. $R_\text{par}$ was varied by changing the resistance of the transistor ($R_\text{tr}$). For the higher temperature (red circles), the data is divided by the temperature ratio $4.2/T_0$. This dataset was acquired during the calibration of heater H1. Based on the assumption that $G$ is independent of the measurement temperature, we slightly adjusted the low bath temperature to a value of $T_0=0.6$~K, which is within 5\% of the actual thermometer reading. The result is shown in Fig.~\ref{FigS3}c. Note that the two power dependencies coincide at small  $R_\text{par}$, where the contribution of the amplifier noise is negligible. At the higher $R_\text{par}$ end, the low temperature data is considerably higher because of the $S_I^\mathrm{Amp}$ contribution. 

When the heater current is turned on, the NW current noise ($S_\text{I}$) increases. The measurement takes place at $V_\text{tr}=0$ V. The measured power modifies according to:
\begin{equation}\label{Pdiode_with_IH}
	P = P_0 +  G \left(  \frac{8k_B T_0}{R_\text{load}} + S_\text{I} + S_\text{I}^\text{Amp}  \right)
\end{equation}

In general, the NW resistance, and thus the current noise gain $G=G(R_\text{par})$, may depend on the $I_\mathrm{H}$. In  order to take this into account, we define the power offset when the heater is turned off:
\begin{equation}\label{Pdiode_with_IH}
	P^* = P_0 +  G^* \left(  \frac{4k_B T_0}{R_\text{par}^*} +  S_\text{I}^\text{Amp}  \right)
\end{equation}
where $G^* = G(R_\text{par}^*)$ and $R_\text{par}^*=R_\text{par}(I_\text{H}=0) $. We extract the NW noise from the power measurement as a function of  $I_\text{H}$:
\begin{widetext}
\begin{equation}
	S_\text{I} = \frac{P-P^*}{G}+\frac{G^*}{G}\frac{4k_BT_0}{R_\mathrm{NW}(I_\mathrm{H}=0)}+\left(\frac{G^*}{G}-1\right)\left(\frac{8k_B T_0}{R_\text{load}} + S_\text{I}^\text{Amp}\right)
\end{equation}
\end{widetext}

In this way we measured the noise power spectral density $S_I$ of the NW and expressed it via the noise temperature $T_N=S_I R_\text{NW}/(4k_B)$.

In the main text, the heater calibration is performed in the normal state, which is achieved by applying a magnetic field $B = 174$~mT. This field fully suppresses superconductivity in the Al island. We observed no systematic difference in the dependence $T_N(I_\mathrm{H})$ when S-island was in the superconducting state ($B=0$), see Figs.~\ref{FigS3}d-e.

\section*{Fitting the 3TM without noise thermometry}

For devices in which a direct Johnson noise calibration was unavailable (D2--D6), the effective bath temperature \(T_{\mathrm{bath}}^{*}(I_{\mathrm{H}})\) was determined via a linear fit based on the known bath and critical temperatures \(T_{\mathrm{c}}(B)\). 

For each device in a fixed magnetic field \(B\), the dependence \(P_{\mathrm{J}}^{\mathrm{c}}(I_{\mathrm{H}})\) was measured. Extrapolation of this dependence to zero yielded the critical heater current \(I_{\mathrm{H}}^{*}\) at which superconductivity is suppressed even in the absence of transport current, i.e. \(P_{\mathrm{J}}^{\mathrm{c}}(I_{\mathrm{H}}^{*})=0\). At this point the effective bath temperature coincides with the superconducting critical temperature: \(T_{\mathrm{bath}}^{*}(I_{\mathrm{H}}^{*}) = T_{\mathrm{c}}(B)\). At \(I_{\mathrm{H}}=0\) the effective bath temperature coincides with the true bath temperature: \(T_{\mathrm{bath}}^{*}(I_{\mathrm{H}}^{*}=0)=T_0\).

We used a linear relationship between \(T_{\mathrm{bath}}^{*}\) and the heater current in the interval \(0 \le I_{\mathrm{H}} \le I_{\mathrm{H}}^{*}\):
\begin{equation}
	T_{\mathrm{bath}}^{*}(I_{\mathrm{H}}) = T_0 + \left( T_{\mathrm{c}}(B) - T_0 \right) \frac{I_{\mathrm{H}}}{I_{\mathrm{H}}^{*}},
	\label{eq:linear_calib}
\end{equation}
which is consistent with the direct calibration by means of noise thermometry in  device D1. Using Eq.~(\ref{eq:linear_calib}), we converted \(I_{\mathrm{H}}\) to \(T_{\mathrm{bath}}^{*}\) and fitted the data with the 3TM. The values of  $\Sigma_\mathrm{env}$  obtained for devices D2--D6 in this way are summarized in Fig.~3e of the main text and in Table~\ref{table}.

\section{Cooling via the InAs core}

\subsection*{Phonon thermal conductivity of the InAs core}


In the main text, we assumed that heat dissipation through the InAs core is negligible. Here we give an upper bound for the heat flux through the core via acoustic phonons in InAs in the ballistic transport regime. We consider a cylindrical InAs NW with cross-sectional area \(S = \pi \rho_{\text{i}}^2\). To avoid explicit summation over quantized transverse modes, we adopt an isotropic phonon gas model, which overestimates the actual heat flux.

The heat flux is calculated analogously to the Stefan–Boltzmann law for acoustic phonons. For a single acoustic branch with an index $\alpha$, the energy flow is written as:
\begin{equation}
\dot{Q}_\alpha = S \int \frac{d^3k}{(2\pi)^3}\, \hbar\omega\, n(\hbar\omega,T)\, v_\alpha \cos\theta,
\end{equation}
where $k$ is the phonon wave vector, $v_\alpha$ is the sound velocity, $n(\hbar\omega,T) = (e^{\hbar\omega/k_B T}-1)^{-1}$ is the Bose–Einstein distribution, and $\theta$ is the angle between the phonon wave vector and the NW axis. The integration is performed over the half-space, which takes into account only phonons propagating in one direction along the NW axis. This  corresponds to half of the total heat flux from the S-island to the reservoirs via InAs. We neglect possible Kapitza resistance between Al and InAs, which, again, overestimates the heat flux. The volume element in \(k\)-space can be expressed as \(d^3k = k^2 \cdot 2\pi \sin\theta \, d\theta \, dk\). Converting the radial variable from \(k\) to \(\omega\) via linear spectrum \(\omega = v_\alpha k\) yields \(d^3k = v_\alpha^{-3} \omega^2 \cdot 2\pi \sin\theta \, d\theta \, d\omega\).
After the integration,
\begin{equation}
\dot{Q}_\alpha = \frac{\pi^2 S (k_B T)^4}{120\hbar^3 v_\alpha^2}.
\end{equation}

Summing over one longitudinal (velocity \(v_L\)) and two transverse (velocity \(v_T\)) modes and multiplying by 2 we obtain the total heat flow from the S-island (\(T_2\)) to the reservoirs (\(T_1\)):
\begin{equation}
\dot{Q}_{ph} = \frac{\pi^2 S k_B^4}{60\hbar^3}\left( \frac{1}{v_L^2} + \frac{2}{v_T^2} \right) \left( T_2^4 - T_1^4 \right).
\end{equation}

Using typical InAs values of the sound velocities \(v_L = 4300\)~m/s, \(v_T = 2400\)~m/s, and taking \(\rho_{\text{i}} = 80\)~nm, we obtain
\begin{equation}
\dot{Q}_{ph} \approx 40\; \frac{\text{pW}}{\text{K}^4} \cdot \bigl(T_2^4 - T_1^4\bigr)
\end{equation}
This value represents the absolute maximum heat current through the InAs core, which can only decrease if the phonon scattering and Kapitza resistance are taken into account.  The critical Joule power measured in experiment is at least an order of magnitude higher, confirming that the dominant cooling pathway is via the surrounding \(^3\)He.




\subsection*{Electronic thermal conductivity of the InAs core}

For completeness, we estimate the electronic heat flow through the InAs core in the diffusive transport regime. The electronic thermal conductivity obeys the Wiedemann–Franz law: \(\kappa_{\text{el}}(T) = L_0 \sigma T\), where \(L_0 = 2.44\times10^{-8}~\text{W}\,\Omega/\text{K}^2\) is the Lorenz number, \(\sigma = L_{\text{InAs}}/(R_{\text{NW}} S)\) is the electrical conductivity, \(L_{\text{InAs}}\) is the NW length, \(R_{\text{NW}}\) is the NW resistance, and \(S\) is the cross-sectional area.

In the diffusive regime, the heat flow along the NW due to a temperature gradient is given by Fourier's law:
\begin{equation}
\dot{Q}_{\text{el}} = -\kappa_{\text{el}}(T) S \frac{dT}{dx},
\end{equation}

Separating variables and integrating from the cold end (\(x=0\), \(T=T_1\)) to the hot end (\(x=L_{\text{InAs}}\), \(T=T_2\)) gives
\begin{equation}
\dot{Q}_{\text{el}} \int_0^{L_{\text{InAs}}} dx = -L_0 \frac{L_{\text{InAs}}}{R_{\text{NW}}} \int_{T_1}^{T_2} T\,dT,
\end{equation}
Cancelling \(L_{\text{InAs}}\) and taking the absolute value (heat flows from hot to cold),
\begin{equation}
\dot{Q}_{\text{el}} = \frac{L_0}{2R_{\text{NW}}}\left(T_2^2 - T_1^2\right).
\end{equation}
The final estimate takes into account that \(R_{\text{NW}} \) is approximately one half of the total NW resistance (\(\sim10~\text{k}\Omega\)) and that the heat flows from the S-island in two directions. Therefore we obtain:
\begin{equation}
\dot{Q}_{\text{el}}\sim  5\frac{\text{pW}}{\text{K}^2}  \cdot \left(T_2^2 - T_1^2\right).
\end{equation}

This is at least two orders of magnitude smaller than the measured critical Joule power, meaning that the contribution of electronic thermal conductance in cooling of the S-island negligible.

\section{ cooling of the S-island via $^3$He gas }

Here we consider the case of cooling a metallic island in a saturated $^3$He vapor. The average mean free path of the $^3$He atom exceeds the size of the S-island and we assume that the atoms are fully thermalized  with the metal surface upon collision. The number flux of atoms incident on the unit surface is $\Phi = \frac{n \bar{v}}{4} $, where \(n\) is the gas concentration and \(\bar{v}=\sqrt{3k_BT_{\text{bath}}/ m_{\text{He}}}\) the root-mean-square speed from the Maxwell–Boltzmann distribution with effective mass $ m_{\text{He}} = 5 \cdot 10^{-27}$ kg. Using the temperature dependence of the saturated vapor pressure \cite{Sydoriak1957}, we determine the atom concentration: $n = P_{\text{He}}/(k_B T_{\text{bath}})$.
Each atom transfers an energy
\begin{equation}
\Delta E = \frac{3}{2} k_B (T_{\text{ph}} - T_{\text{bath}})
\end{equation}
upon equilibration with the S-island at temperature \(T_{\text{ph}}\). The heat flux per unit area becomes $\frac{Q}{A} = \Phi \Delta E$.
Comparing with the 3TM expression $\frac{Q}{A} = G_{\text{env}}(T_{\text{ph}}-T_{\text{bath}})$ gives the environmental thermal conductance
\begin{equation}
G_{\text{env}} = \frac{3}{8} P_{\text{He}} \sqrt{\frac{3k_B}{m_{\text{He}} T_{\text{bath}}}} .
\end{equation}

\section{ cooling of the S-island via liquid $^3$He}

We consider a cylindrical S-island of radius \(r_0 = \rho_{\text{i}} + t\) and length \(L \gg r_0\), with its surface (phonon) temperature of \(T_0\). The island is embedded in liquid \(^3\)He, whose thermal conductivity depends on temperature as \(\kappa(T) = \alpha T^m\). Our goal is to find the total heat flux \(\dot{Q}\) from the island to the bath at temperature \(T_{\text{bath}}\) (with \(T_0 > T_{\text{bath}}\)).

We consider a stationary heat conduction equation without internal heat sources in cylindrical coordinates, which defines the heat flux divergence: $ \mathbf{\nabla} \cdot  \mathbf{q} = 0 $. The heat flux density \(q\) (power per unit area) obeys Fourier's law: \(q = -\kappa(T)\, dT/dr\), where \(r\) is the radial coordinate. The total heat flux through a surface of area \(S\) is \(\dot{Q} = S q\) (the sign of \(q\) indicates the direction).

To analyze the heat transfer, we split space around the S-island into two regions. Close to the cylinder (Region I, \(r_0 \le r \le R\)), the temperature gradient is predominantly radial because \(L \gg r_0\); the heat flow is governed by cylindrical symmetry. Far from the cylinder (Region II, \(r \ge R\)), the heat flow becomes approximately isotropic and spherically symmetric. The matching radius \(R\) is taken to be of order \(L\); its optimal value is determined below.

\paragraph{Region I (cylindrical symmetry).}
For a cylindrical surface of radius \(r\) and length \(L\), the total heat flux is \(\dot{Q} = -2\pi r L \alpha T^m (dT/dr)\). Separating variables and integrating from the island surface (\(r = r_0\), \(T = T_0\)) to the matching radius (\(r = R\), \(T = T(R)\)) yields:
\begin{equation}
	T_0^{m+1} - T(R)^{m+1} = \frac{\dot{Q}(m+1)}{2\pi\alpha L} \ln\left(\frac{R}{r_0}\right). 
\end{equation}

\paragraph{Region II (spherical symmetry).}
For a spherical surface of radius \(r\), the total heat flux is \(\dot{Q} = -4\pi r^2 \alpha T^{m} (dT/dr) \). Proceeding similarly, we integrate from \(r = R\) (\(T = T(R)\)) to infinity (\(T = T_{\text{bath}}\)):
\begin{equation}
	T(R)^{m+1} - T_{\text{bath}}^{m+1} = \frac{\dot{Q}(m+1)}{4\pi\alpha} \frac{1}{R}. 
\end{equation}

\paragraph{Matching the two regions.}
Adding equations (6) and (7) eliminates the unknown intermediate temperature at the matching radius \(T(R)\):
\begin{equation}
	T_0^{m+1} - T_{\text{bath}}^{m+1} = \frac{\dot{Q}(m+1)}{2\pi\alpha L} \ln\left(\frac{R}{r_0}\right) + \frac{\dot{Q}(m+1)}{4\pi\alpha} \frac{1}{R}.
\end{equation}
Solving for \(\dot{Q}\) gives:
\begin{equation}
	\dot{Q} = \frac{2\pi\alpha L}{m+1} \left[ \ln\left(\frac{R}{r_0}\right) + \frac{L}{2R} \right]^{-1} \left( T_0^{m+1} - T_{\text{bath}}^{m+1} \right).
\end{equation}

The matching radius \(R\) is determined from the condition of maximum cooling power.  We find the minimum of the bracket in the denominator with respect to \(R\) by differentiation, which yields the optimal value \(R = L/2\). Substituting this back gives the final expression:
\begin{equation}\label{dotQ_fin}
	\dot{Q} = \frac{2\pi\alpha L}{m+1} \cdot \frac{1}{\ln\left(\dfrac{L}{2r_0}\right) + 1} \cdot \left( T_0^{m+1} - T_{\text{bath}}^{m+1} \right). 
\end{equation}
Eq.~\eqref{dotQ_fin} provides the effective environmental cooling power $\Sigma_{\text{env}}$ used in the three-temperature model (Eq.~(1) of the main text). For liquid $^3$He with thermal conductivity $\kappa(T) = \alpha T^{0.5}$ ($m = 0.5$). We take into account that the liquid occupies only the half-space above the substrate, which reduces the above estimate by two and yields $\Sigma_{\text{env}}$ values plotted in Fig.~3e of the main text (dashed line). This model overestimates the experimental value of $\Sigma_{\text{env}}$ by approximately a factor of 10. This systematic discrepancy suggests that the heat is not dissipated into a bulk half-space of liquid $^3$He, but rather into a thin wetting film. 


\section{non-reproducible critical power behavior}

\begin{figure}[t!]
	\begin{center}
		\vspace{0mm}
		\includegraphics[width=1\linewidth]{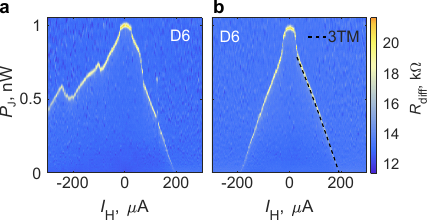}
	\end{center}
	\caption{ Color-scale plot of the $R_\text{diff} = dV/dI_\text{NW}$ as a function of Joule power $P_\text{J}$ in the NW and $I_\text{H}$ for the device D6.
		The dashed line represents the fit using the three-temperature model (3TM). } 
	\label{FigS4}
\end{figure}

The dependencies of the critical power on the heater current presented in the main text are obtained in the steady-state regime. The non-steady-state regime refers to a state where the critical Joule power $P_\mathrm{J}^\mathrm{c}$  varies randomly and irreproducibly with time. After cooling the device chamber from 4.2~K down to the operating temperatures, approximately one hour is required to reach the steady-state regime. We attribute this behavior to the peculiarities of \(^3\)He cooling and the possible formation of bubbles and boiling of \(^3\)He near heaters. Figure~\ref{FigS4} illustrates the approach to the steady-state regime for device D6, which exhibits a tiny hysteresis of the $I$-$V$ curves near the superconducting-normal transition in the S-island. Two consecutive maps of the differential resistance versus Joule power and heater current \(I_{\text{H}}\) are shown in Fig.~\ref{FigS4}a and Fig.~\ref{FigS4}b.  In Fig.~\ref{FigS4}a, the dependence $P_\mathrm{J}^\mathrm{c}$ on the $|I_\mathrm{H}|$ is asymmetric, non-monotonic and irreproducible. This is a non-steady-state regime, which we attribute to random variations of the $\Sigma_{\text{env}}$. Note that the heater current is swept from negative to positive values and the behavior at $I_{\text{H}}>0$ is more regular. By contrast, in Fig.~\ref{FigS4}b the behavior is stable, allowing a fit based on the three-temperature model (3TM), which is shown by the dashed line. This fit is used in Fig.~3e of the main text.

In the regime of small heater currents, a distinct reproducible feature is observed in Fig.~\ref{FigS4}b. Here, the critical Joule power shows a local flat maximum that is not captured by the 3TM with the constant $\Sigma_{\text{env}}$. 
This feature is common to all devices and is reproducibly observed at low enough bath temperatures, as also visible in Fig.~1d of the main text. 
The feature disappears above approximately $T_\mathrm{bath}\approx$0.4~K, that is why it is absent in Figs.~2c and~3a of the main text. We attribute this behavior to a change (decrease) of the environmental cooling parameter $\Sigma_{\text{env}}$ at increasing the heater current. Although the microscopic details of this phenomenon are not understood, a plausible explanation is that the thickness of the $^3$He layer wetting the NW decreases at increasing $I_\mathrm{H}$.

\bibliography{biblio.bib}